\documentclass[useAMS,usenatbib,pdftex,usegraphicx]{mn2e}

\newcommand{\kms}{km s$^{-1}$}
\newcommand{\cmN}{cm$^{-2}$}

\newcommand{\lam}{$\lambda$}
\newcommand{\civ}{\mbox{C\,{\sc iv}}}

\newcommand{\siiv}{\mbox{Si\,{\sc iv}}}

\newcommand{\siv}{\mbox{S\,{\sc iv}}}
\newcommand{\svi}{\mbox{S\,{\sc vi}}}

\newcommand{\ovi}{\mbox{O\,{\sc vi}}}

\newcommand{\al}{\mbox{Al\,{\sc iii}}}
\newcommand{\mgii}{\ifmmode {\rm Mg}{\textsc{ii}} \else Mg\,{\sc ii}\fi}
\newcommand{\heii}{\ifmmode {\rm He}{\textsc{ii}} \else He\,{\sc ii}\fi}
\newcommand{\pv}{\mbox{P\,{\sc v}}}

\title[\pv\ absorption in BOSS]
{Quasars with \pv\ broad absorption in BOSS data release 9}
\author[D. M. Capellupo et al.]{D. M. Capellupo$^{1,2}$
\thanks{E-mail:danielc@physics.mcgill.ca (DMC)},
F. Hamann$^{3,4}$, H. Herbst$^{4}$,
W. N. Brandt$^{5,6,7}$,
J. Ge$^{4}$,\newauthor
I. P\^aris$^{8}$,
P. Petitjean$^{9}$,
D. P. Schneider$^{5,6}$,
A. Streblyanska$^{10,11}$,
D. York$^{12,13}$ \\
$^{1}$Department of Physics, McGill University, Montreal, Quebec, H3A 2T8, Canada \\
$^{2}$McGill Space Institute, McGill University, Montreal, QC, H3A 2A7, Canada \\
$^{3}$Department of Physics and Astronomy, University of California, Riverside, CA 92507, USA \\
$^{4}$Department of Astronomy, University of Florida, Gainesville, FL 32611-2055, USA \\
$^{5}$Department of Astronomy and Astrophysics, The Pennsylvania State University, University Park, PA 16802 \\
$^{6}$Institute for Gravitation and the Cosmos, The Pennsylvania State University, University Park, PA 16802 \\
$^{7}$Department of Physics, Pennsylvania State University, University Park, PA 16802, USA \\
$^{8}$Aix Marseille Universit\'e, CNRS, LAM, (Laboratoire d'Astrophysique de Marseille), Marseille, France \\
$^{9}$Institut d'Astrophysique de Paris, Universit\'{e} Paris 6-CNRS, UMR7095, 98bis Boulevard Arago, F-75014 Paris, France \\
$^{10}$Instituto de Astrofisica de Canarias (IAC), E-38200 La Laguna, Tenerife, Spain \\
$^{11}$Departamento de Astrofisica, Universidad de La Laguna (ULL), E-38205 La Laguna, Tenerife, Spain \\
$^{12}$Department of Astronomy \& Astrophysics, University of Chicago, Chicago, IL 60637 \\
$^{13}$The Enrico Fermi Institute, University of Chicago, Chicago, IL 60637
}

\begin{document}


\pagerange{\pageref{firstpage}--\pageref{lastpage}} \pubyear{2002}

\maketitle

\label{firstpage}

\begin{abstract}
Broad absorption lines (BALs) found in a significant fraction of quasar spectra
identify high-velocity outflows that might be present in all quasars and could
be a major factor in feedback to galaxy evolution. Understanding the nature of
these flows requires further constraints on their physical properties,
including their column densities, for which well-studied BALs, such as
\civ\ \lam\lam1548,1551, typically provide only a lower limit because of
saturation effects. Low-abundance lines, such as \pv\ \lam\lam1118,1128,
indicate large column densities, implying outflows more powerful than
measurements of \civ\ alone would indicate. We search through a sample of 2694
BAL quasars from the SDSS-III/BOSS DR9 quasar catalog for such absorption, and
we identify 81 `definite' and 86 `probable' detections of \pv\ broad
absorption, yielding a firm lower limit of 3.0$-$6.2\% for the incidence of
such absorption among BAL quasars. The \pv-detected quasars tend to have
stronger \civ\ and \siiv\ absorption, as well as a higher incidence of LoBAL
absorption, than the overall BAL quasar population. Many of the \pv-detected
quasars have \civ\ troughs that do not reach zero intensity (at velocities
where \pv\ is detected), confirming that the outflow gas only partially covers
the UV continuum source.
\pv\ appears significantly in a composite spectrum of non-\pv-detected BAL
quasars, indicating that \pv\ absorption (and large column densities) are much
more common than indicated by our search results. Our sample of \pv\ detections
significantly increases the number of known \pv\ detections, providing
opportunities for follow-up studies to better understand BAL outflow
energetics.
\end{abstract}


\section{Introduction}
\label{sec:intro}

High velocity outflows, originating from quasar accretion discs, likely exist
in all quasars and may be an important contributor to feedback to galaxy
evolution. One of the signatures of these outflows is broad absorption lines
(BALs) in quasar spectra. While numerous works have studied BALs, many
properties of the outflows themselves are still uncertain. Further information
on their location, column densities, mass outflow rates, and energetics is
required to understand the nature of these flows.

In some cases, the distances of the flows from the central black hole can be
inferred from BAL variability
\citep{Misawa07,Moe09,Capellupo11,Capellupo13,Capellupo14,Hall11,RodriguezH11,RodriguezH13}
or from excited state lines with photoionization modelling
\citep{Moe09,Dunn10,Borguet13}.

Column density estimates, in general, provide only lower limits because of
saturation. In an unknown number of cases, measuring the apparent optical depth
of a BAL trough directly from the observed spectra provides only a lower limit
on the true optical depth in the line, and thus the column density of the flow,
because the absorbing gas only partially covers the background light source
\citep[e.g.][]{Hamann98,Arav99a,Gabel03}. Constraining the energetics of the
flows requires knowledge of both the locations and the column densities of the
outflows.

One method that overcomes this problem of saturation in BALs is searching for
absorption in ions of low abundance. \pv\ is such an ion, for P/C $\sim$ 0.001
in the Sun \citep{Asplund09}. Besides its low abundance, \pv\ is a good choice
because it has a resonance doublet at wavelengths 1118 and 1128 \AA, easily
accessible from ground-based observations of quasars of moderate to high
redshift, and its ionization is similar to much more abundant and commonly
measured ions such as \civ\ \lam\lam1548,1551. \pv\ absorption should be
present if the column densities in the outflows are sufficiently large.
Furthermore, if the relative abundances are similar to solar abundances,
\pv\ absorption implies that commonly observed BALs, such as \civ, are actually
highly optically thick and the total outflow column densities are much larger
than the apparent optical depths would indicate.

BAL quasars are typically identified based on the presence of \civ\ broad absorption. Several studies have used the assumption of solar abundances and a standard ionizing spectrum in photoionization models to determine the true \civ\ optical depths, as well as the true total column densities, based on the presence of \pv\ absorption. \citet{Hamann98} finds that the true \civ\ optical depths are at least $\sim$800 times greater than the \pv\ optical depth in idealized BAL clouds that are optically thin throughout the Lyman continuum. The ratio of \civ\ to \pv\ optical depths might be as low as $\sim$100 in other situations with total column densities up to $N_H \sim 4\times 10^{23}$ cm$^{-2}$. Similar results, across a wide range of physical conditions, are found by \citet{Leighly09,Leighly11}. \citet{Borguet12} use many more observational constraints on the outflow in a particular quasar to determine a \civ/\pv\ optical depth ratio of $\sim$1200. Therefore, the presence of even weak \pv\ absorption indicates a highly saturated \civ\ BAL and, therefore, large total column densities.

Building upon the results of \citet{Hamann98} and \citet{Leighly09,Leighly11},
\citet{Capellupo14} estimate that the detection of a strong \pv\ BAL, with
apparent optical depth reaching as high as 1.5, indicates
log $N_H$ $>$ 22.3 \cmN. Furthermore, they argue that the detection of
variability in a saturated \civ\ trough, at the same velocities as the \pv\
absorption, strongly favors the hypothesis of outflow clouds crossing our line
of sight to the continuum source. This behaviour indicates distances much
smaller than those determined via excited state line analysis (sub-parsec to pc
scales versus $\sim$kpc scales; \citealt{Chamberlain15} and references
therein). Despite the smaller distance, \citet{Capellupo14} estimate the ratio
of the kinetic energy luminosity of the flow to the bolometric luminosity of
the quasar to be of the order of magnitude necessary for feedback to the host
galaxy (0.005 to 0.05; e.g. \citealt{Prochaska09,Hopkins10}).

Altogether, these previous results demonstrate the importance of the detection
of low abundance ions in understanding the energetics of BAL outflows. To date,
however, detections of \pv\ absorption have been limited to a handful of
individual cases
\citep{Turnshek88,Junkkarinen97,Hamann98,Borguet12,Capellupo14}.
Identifying clear detections of \pv\ absorption in a large sample of BAL
quasars would allow comparisons between BAL quasars with such detections and
those without. Certain sub-groups of BAL quasars, namely LoBALs and FeLoBALs,
are known to have different properties than the overall BAL population. For
example, they tend to have redder colours than the overall BAL quasar population
\citep[e.g.,][]{Gibson09}. \citet{FilizAk14} finds evidence for a correlation
between the incidence of \pv\ absorption and the existence of \siiv\ and \al\
absorption.
It will be useful for the understanding of BAL quasars in general to further
investigate how quasars with \pv\ detections, which are indicative of large
outflow column densities, are similar or different to BALs without \pv\
absorption.
Furthermore, a large list of \pv\ detections will
prove useful for future studies at other wavelengths that can further assess
the energetics and other properties of quasars with \pv\ detections as compared
to those without.

Therefore, we perform a systematic search for \pv\ absorption in the Sloan
Digital Sky Survey III (SDSS-III; \citealt{Eisenstein11}) Baryon Oscillation
Spectroscopic Survey (BOSS; \citealt{Dawson13}), which uses the SDSS 2.5-meter
telescope at Apache Point Observatory \citep{Gunn06}. The BOSS survey is
ideally suited to this search as it was designed to target over 100,000 quasars
at redshifts $z_e > 2.2$ \citep{Bolton12,Ross12}, and, as a result, it contains
one of the largest samples of BAL quasars. Compared to the original SDSS-I/II
survey \citep{York00}, the BOSS spectrograph \citep{Smee13} has smaller fibers,
improved detectors, higher throughput, and a wider wavelength range, extending
to $\sim$3600 \AA\ at the blue end \citep{Dawson13}.
Despite the more favorable wavelength coverage, our search is limited to
quasars at a redshift $z_e > 2.3$ because of the location of the \pv\ doublet.

Furthermore, since the wavelength of \pv\ places it in the Ly$\alpha$ forest,
it is nearly impossible to rule out the existence of weak or narrow \pv.
Therefore, our systematic search will only be able to identify those BAL
quasars with a clear detection of \pv, at the same outflow velocities as the
\civ\ BALs, which were originally used to identify BAL quasars among the BOSS
quasar sample.

Section~\ref{sec:sample} describes our parent sample of BAL quasars and our
method of searching for \pv\ absorption. Section~\ref{sec:res} discusses the
general properties of the quasars with \pv\ detections. In
Section~\ref{sec:comp}, we present composite spectra of BAL quasars both with
and without \pv\ detections, and we conclude with some discussion in
Section~\ref{sec:discuss}.

\section{Sample Selection and \pv\ Search}
\label{sec:sample}

The SDSS-III BOSS DR9 \citep{Ahn12} quasar (DR9Q) catalog \citep{Paris12}
includes measurements of the balnicity index (BI; \citealt{Weymann91}), which
is the standard indicator of a BAL quasar.
In order to cover the full range of BAL strengths, we select those quasars
with BI $>$ 0 and BI $> 3\sigma_{BI}$. Because of the wavelengths of the
\pv\ doublet, 1118 and 1128 \AA, we must limit our sample to $z_e > 2.3$.
The BOSS spectrograph provides coverage down to 3600 \AA, which corresponds
to 1090 \AA\ in the rest-frame of a $z_e$ $\sim$ 2.3 object.
These selection criteria produce a parent sample of 2694 BAL quasars.

Two of the authors, DMC and FH, independently visually inspected all 2694 of
the DR9Q BAL quasars that met our BI criteria to search for detections of \pv\
absorption. We examine both the full spectrum and the individual wavelength
regions of the \civ\ absorption and potential \siiv\ and \pv\ absorption. We
used the velocities of the \civ\ and, if it is present, \siiv\ troughs to guide
our search for \pv. The location of \pv\ in the Ly$\alpha$ forest means that
any potential \pv\ absorption is likely to be blended with unrelated,
intervening absorption. Another difficulty, particularly for those quasars near
the lower redshift cutoff, is the decrease in the signal-to-noise ratio towards
the blue end of the spectrum.

\begin{figure*}
  \centering
  \includegraphics[width=150mm]{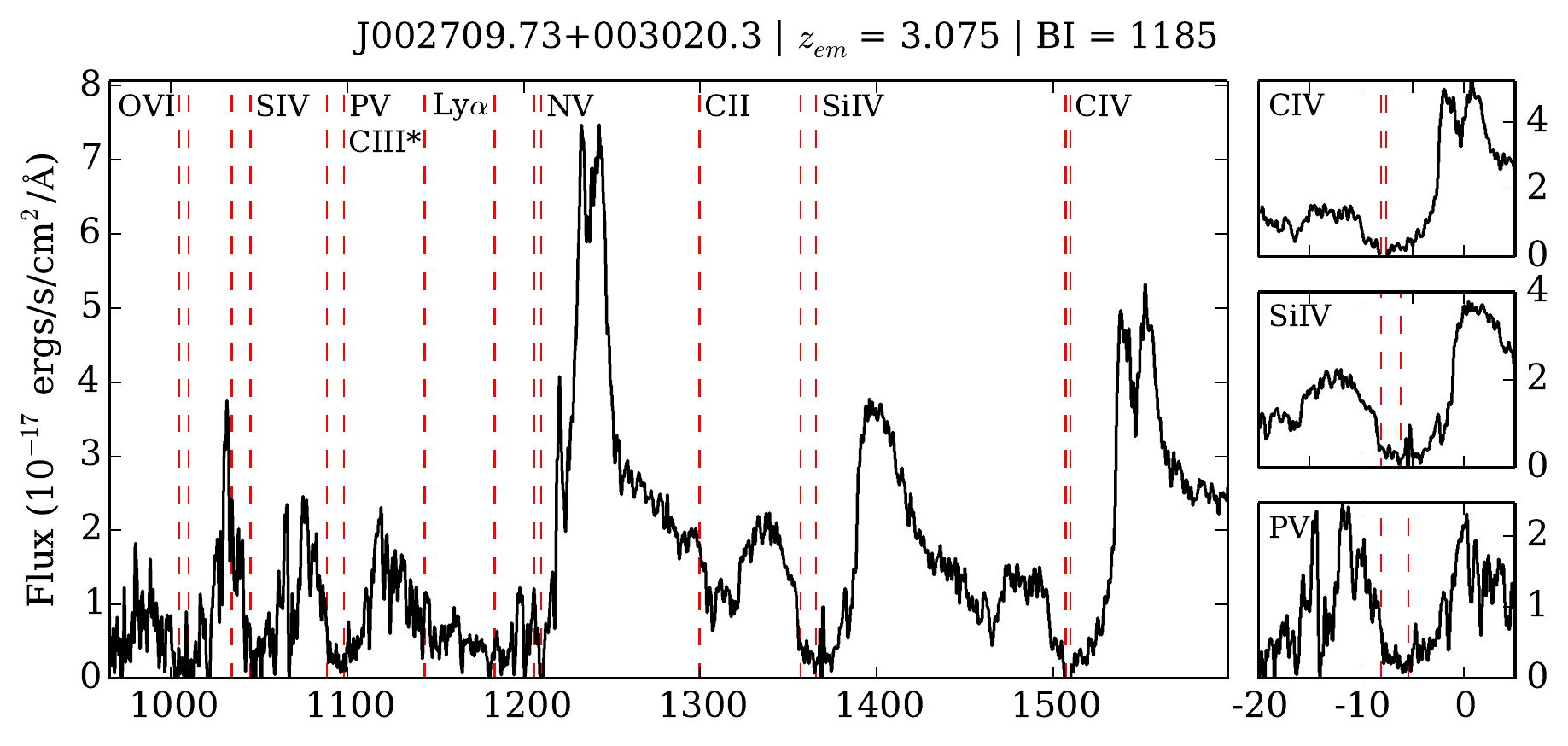}
  \includegraphics[width=154mm]{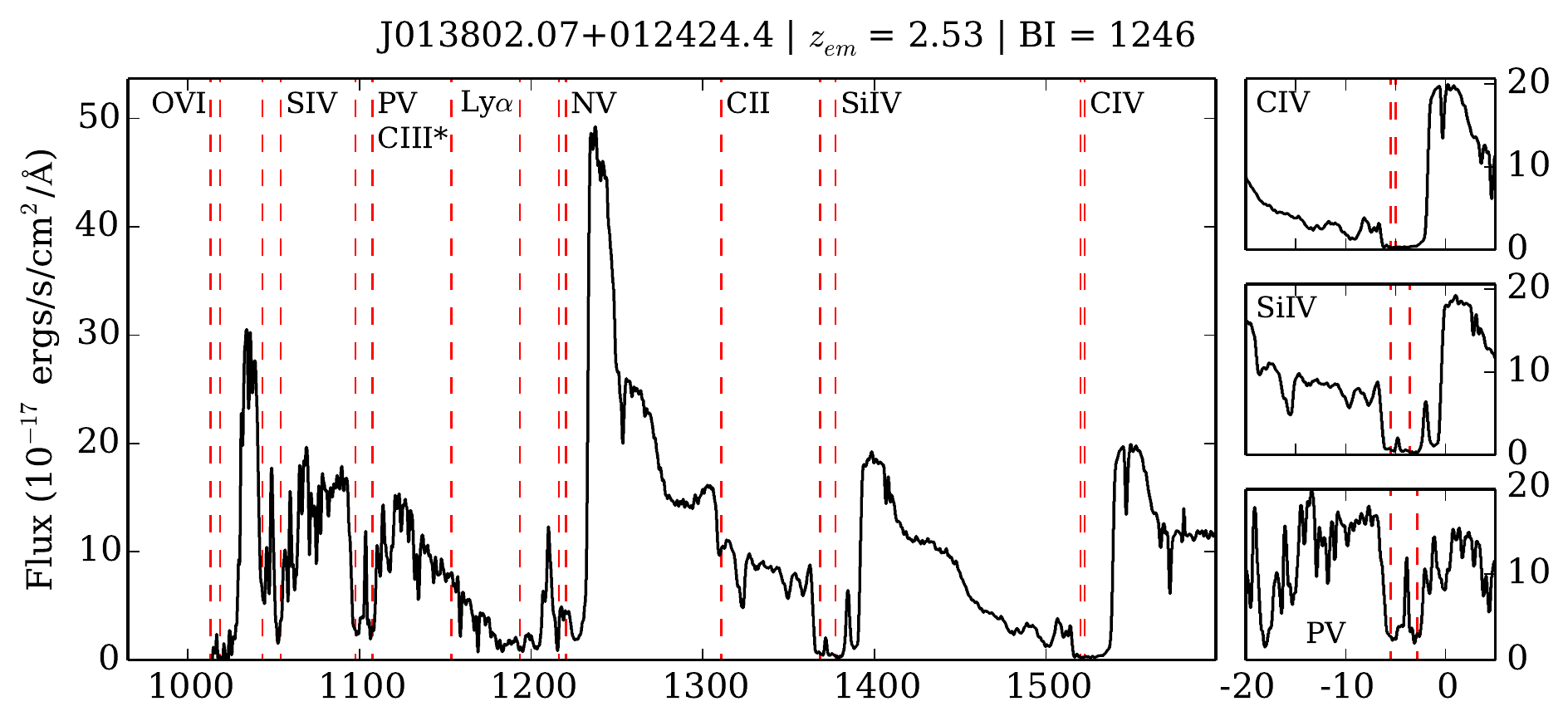}
  \includegraphics[width=154mm]{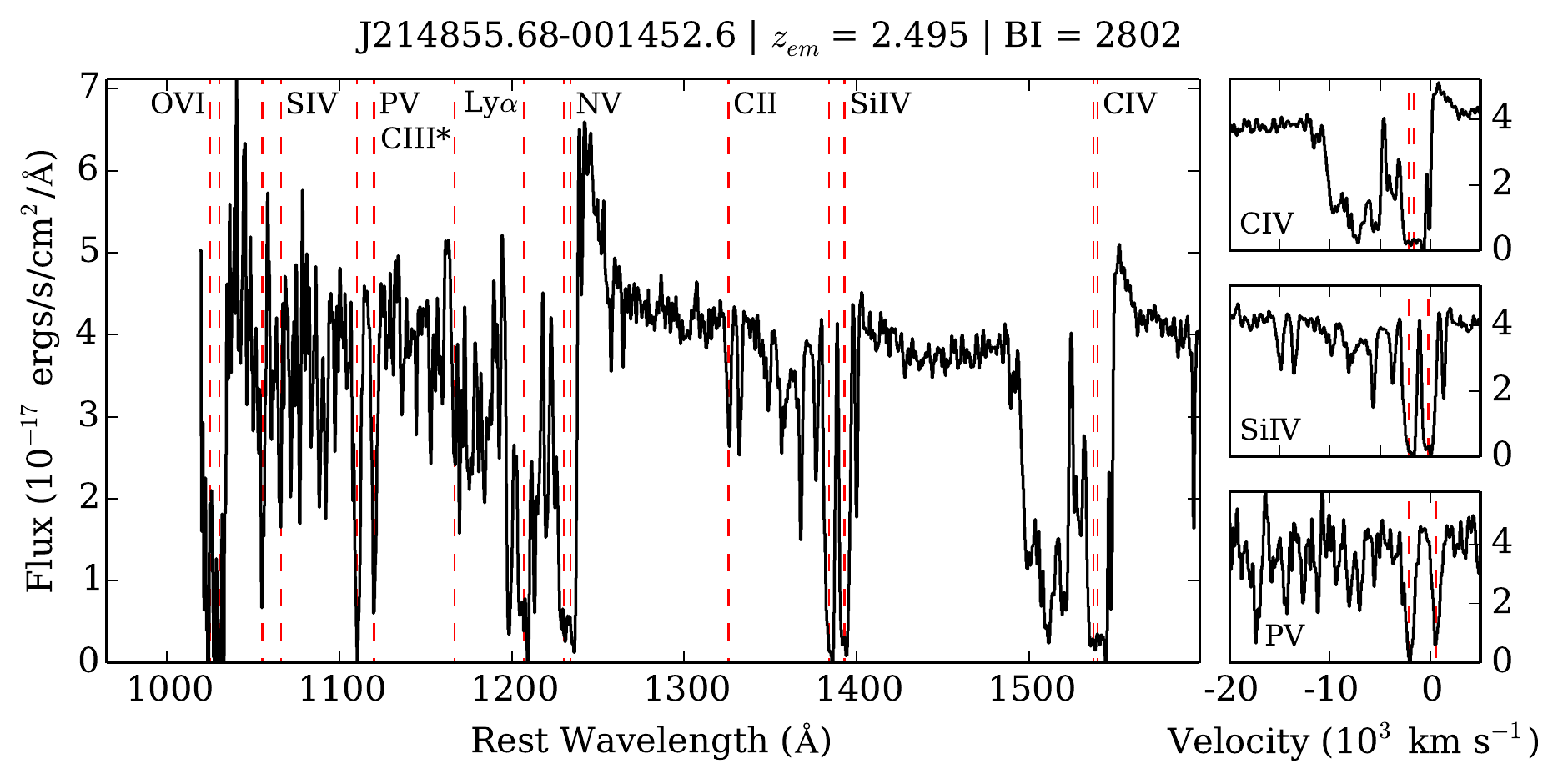}
 \caption{Examples of three of the 81 `definite' detections of \pv\ absorption,
    with varying \pv\ profile shapes. For each quasar, the full spectrum from
    blueward of \ovi\ to redward of \civ\ is displayed in the left panel; in
    the right panel, only the \civ, \siiv, and \pv\ absorption regions are
    plotted on a velocity scale. The potential locations of various absorption
    lines are labeled based on the velocity of the deepest segment of the \civ\
    trough.
    Spectra of all \pv-detected quasars are available online.}
 \label{fig:defspec}
\end{figure*}

\begin{figure*}
  \centering
  \includegraphics[width=150mm]{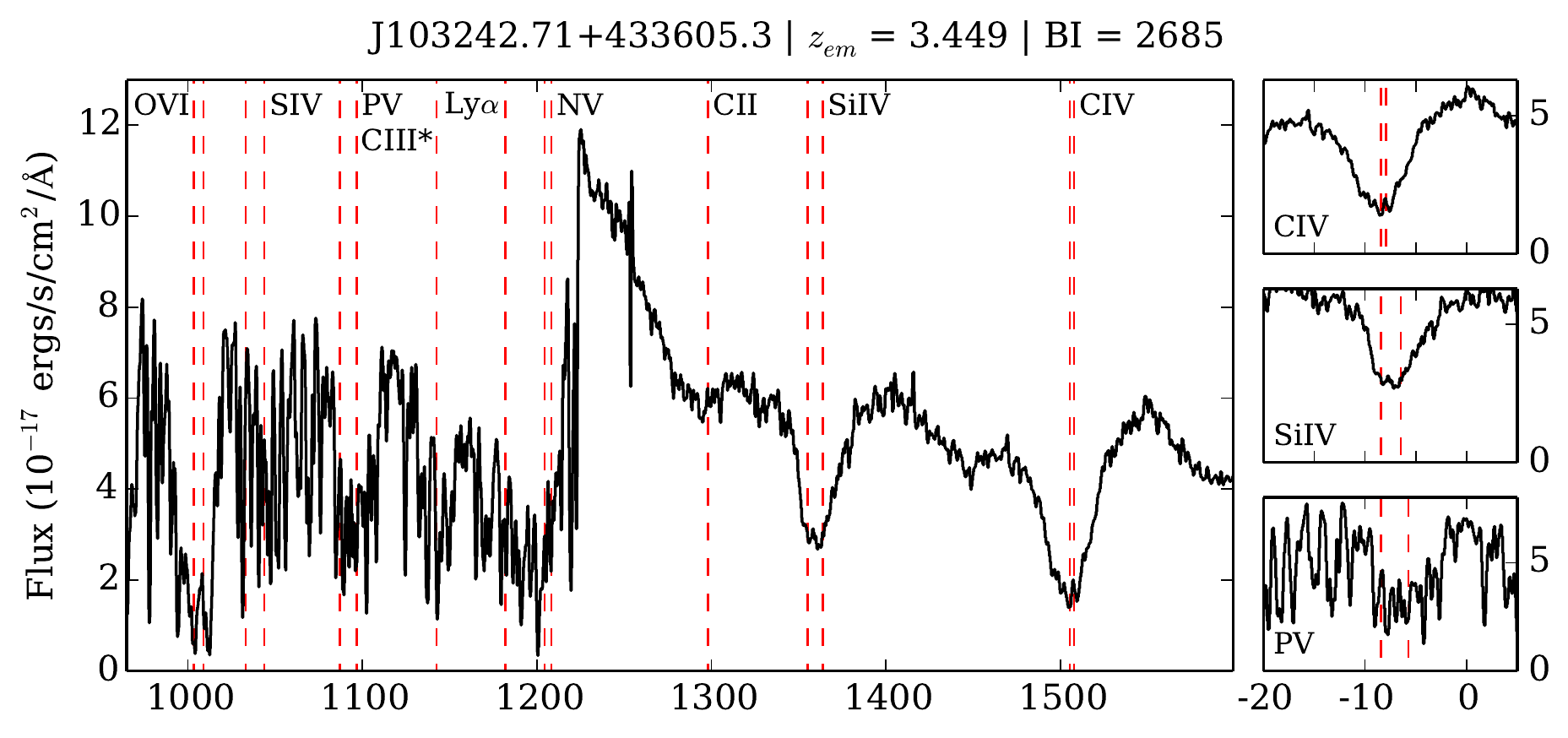}
  \includegraphics[width=150mm]{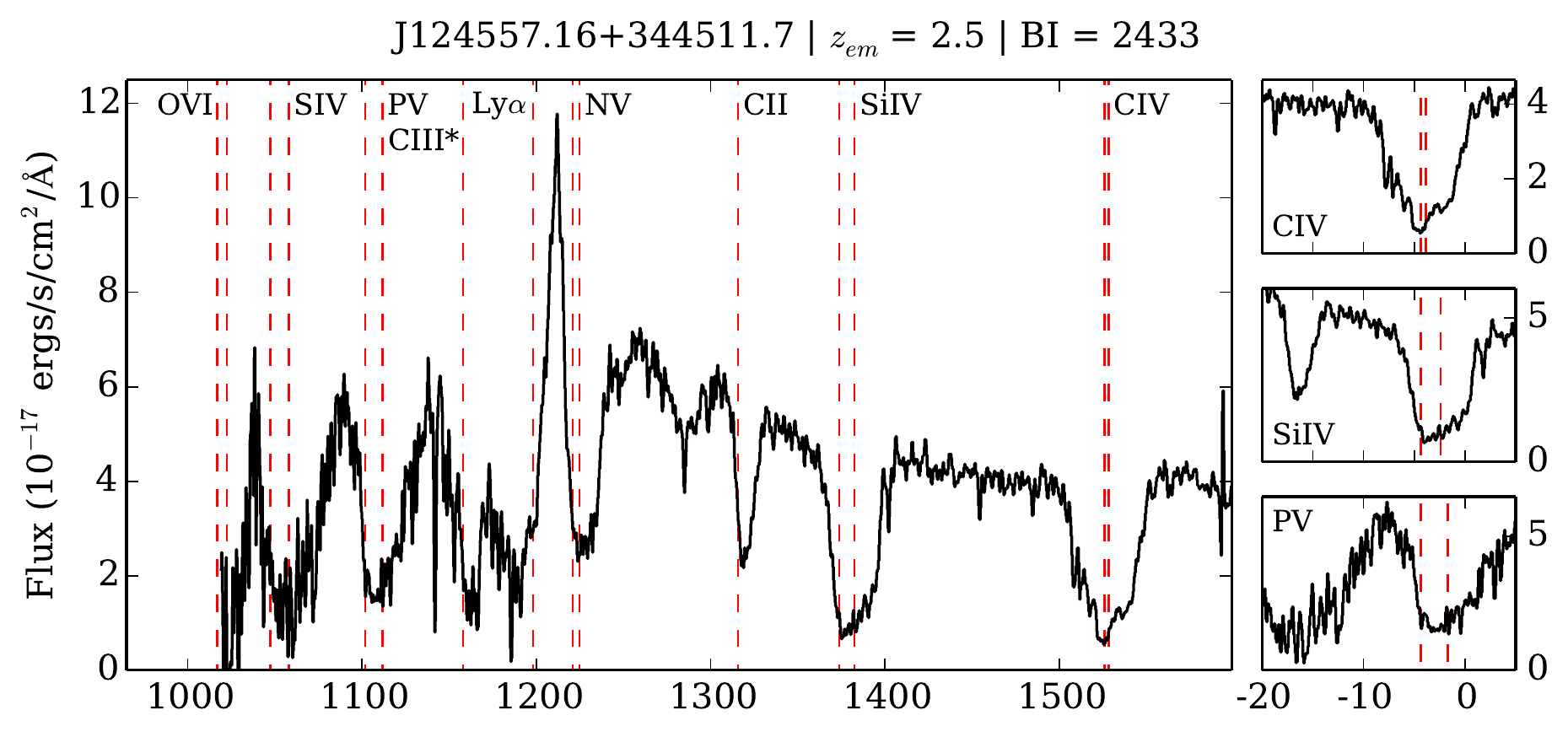}
  \includegraphics[width=150mm]{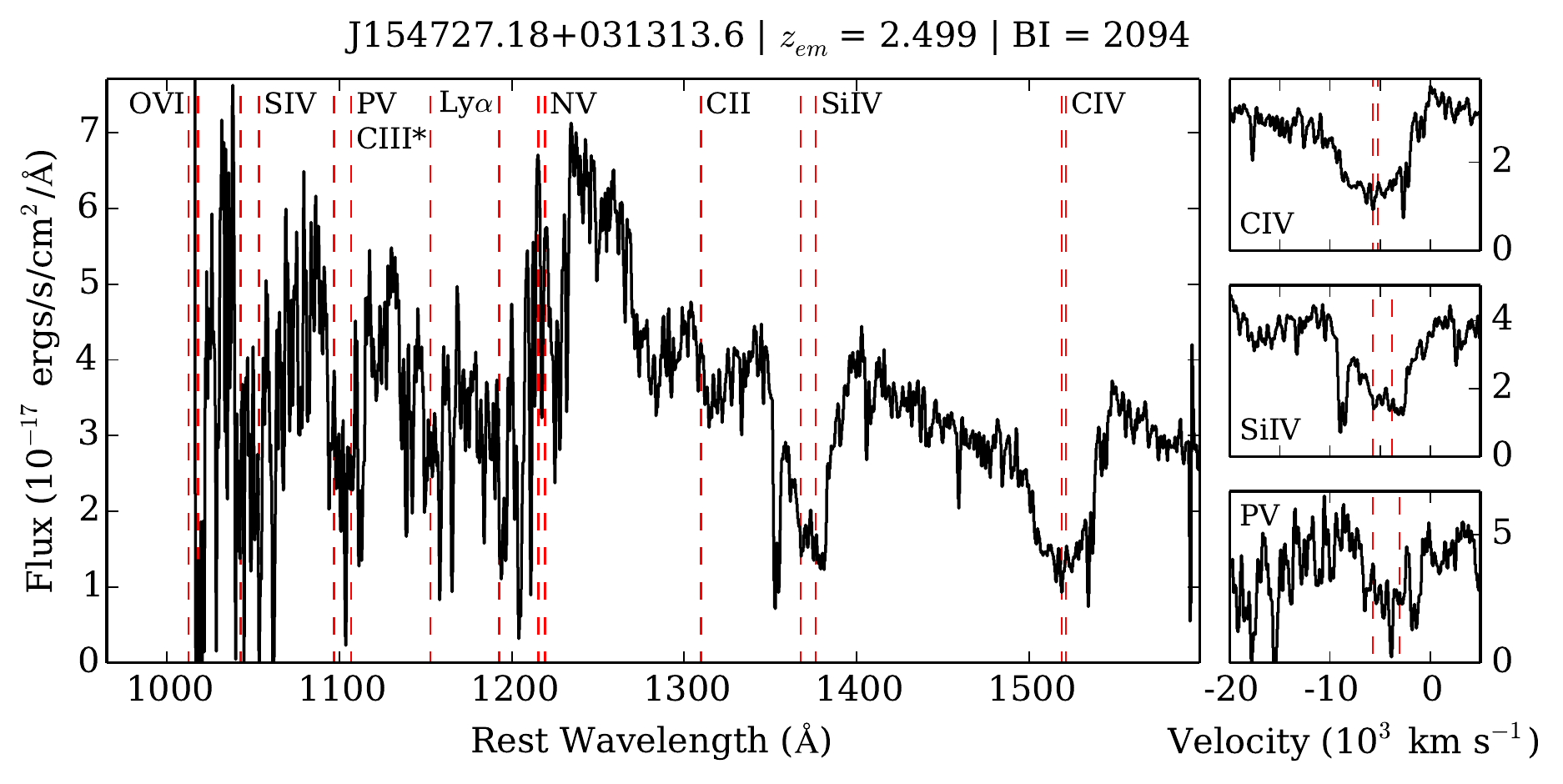}
  \caption{Same format as Fig. \ref{fig:defspec}, but with three examples with
    varying \civ\ depth.}
  \label{fig:defspec2}
\end{figure*}

\begin{figure*}
  \centering
  \includegraphics[width=150mm]{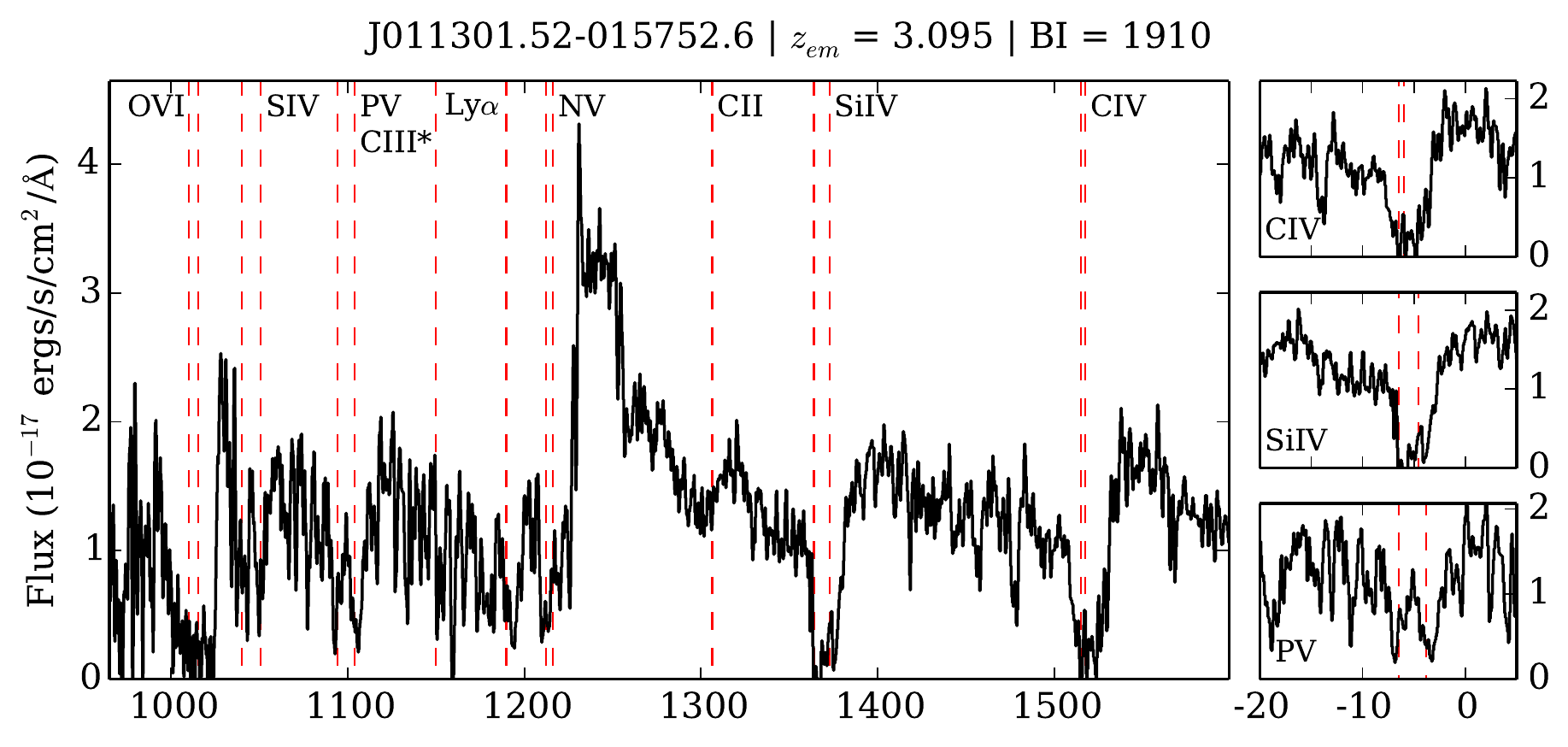}
  \includegraphics[width=150mm]{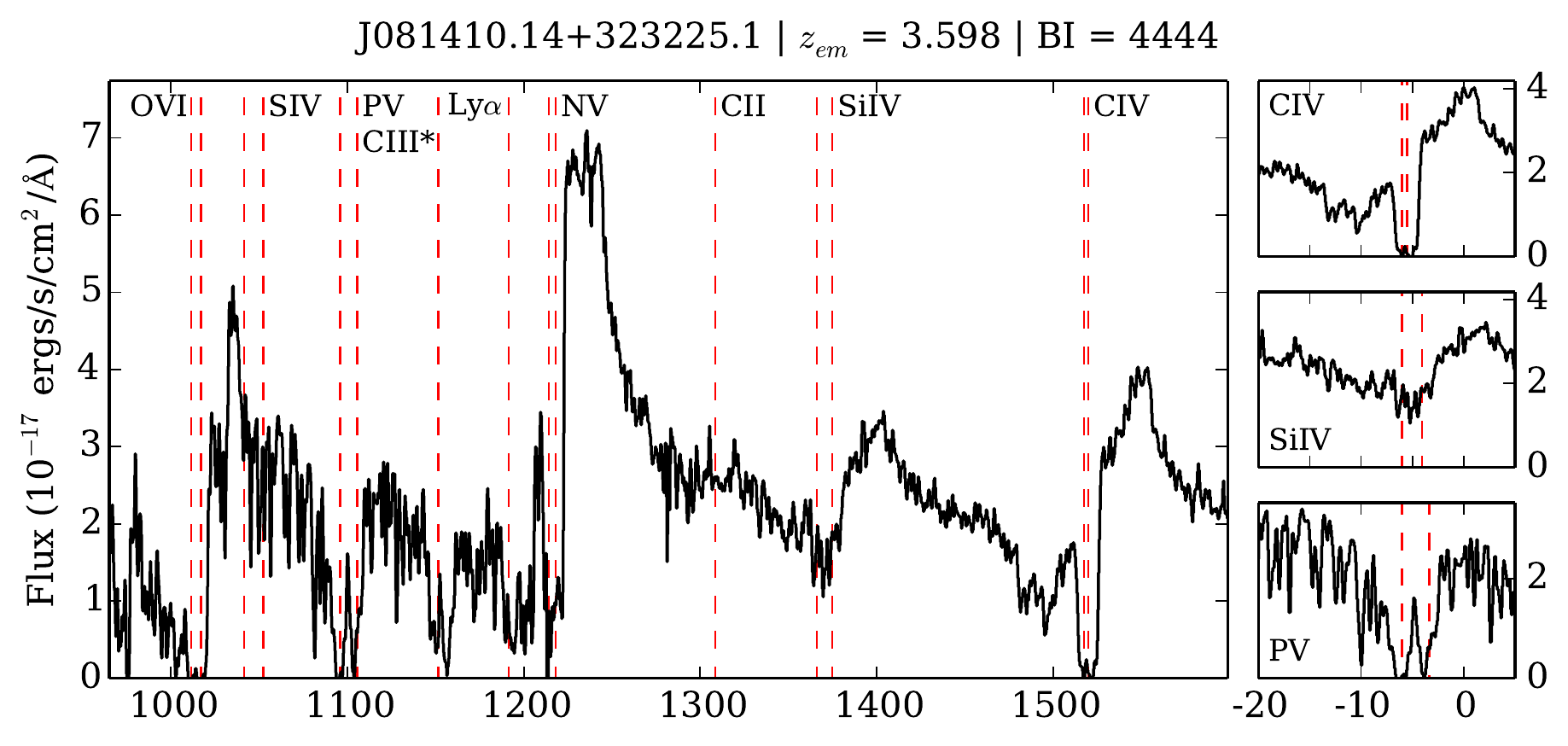}
  \includegraphics[width=155mm]{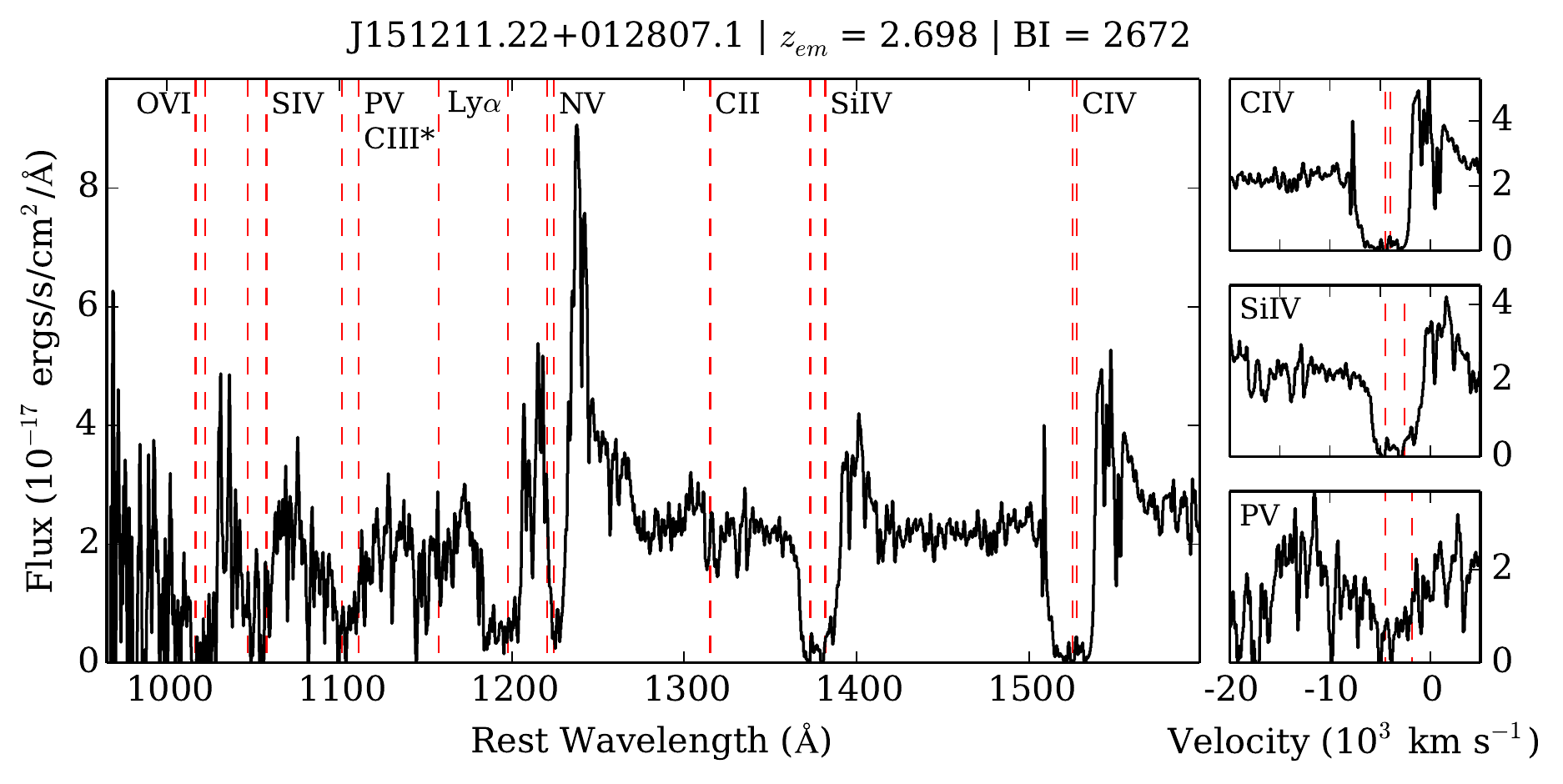}
  \caption{Same format as Fig. \ref{fig:defspec}, but for three of the 86
    `probable' \pv\ detections.}
  \label{fig:probspec}
\end{figure*}

For these reasons, we divide our candidate \pv\ detections into two lists:
`definite' and `probable'. Figs \ref{fig:defspec} to \ref{fig:probspec} show
nine example spectra of these `definite' and `probable' \pv-detected quasars
(spectra of all the \pv-detected quasars are available online).
For each quasar, we plot the rest wavelength range 965 to 1600 \AA, which
includes lines from \ovi\ $\lambda$1037 through \civ. To the right of each spectrum are inset
plots of the \civ, \siiv, and \pv\ BALs. Broad, deep absorption at the same
velocity, with similar velocity width and profile shape, as \civ\ and/or \siiv,
in a relatively high signal-to-noise ratio spectrum, indicates a secure
`definite' detection of \pv\ (for example, J002709+003020 in
Fig.~\ref{fig:defspec} and J124557+344511 in Fig.~\ref{fig:defspec2}). If the
signal-to-noise ratio is low, then we prefer to include such a detection in the
`probable' list, in order to be conservative in our classifications.
In some cases, there is doublet structure at the expected
location, matching the expected wavelength separation, of \pv, providing a
strong confirmation that it is a `definite' detection of \pv, such as in
J013802.07+012424.4 and J214855.68-001452.6 in Fig.~1. There appears to be a
doublet at the expected location of \pv\ in J011301.52-015752.6 and
J081410.14+323225.1 in Fig. \ref{fig:probspec}, but the relatively low
signal-to-noise ratio in J011301.52-015752.6 and the weak \siiv\ in
J081410.14+323225.1 makes these `probable' detections, rather than secure
`definite' detections.

The location of \pv\ in the Ly$\alpha$ forest and our reliance on \civ\ and/or
\siiv\ absorption to guide our search introduces several biases. If the \pv\
absorption is narrower than the \civ\ and/or \siiv\ absorption, it can be
easily confused with Ly$\alpha$ forest lines, leading to a bias against these
narrow \pv\ sources.
Furthermore, previous work on individual cases of \pv\ absorption have found
that the shape of the \siiv\ trough is a better match to the shape of the \pv\
absorption than is \civ\ \citep{Junkkarinen97,Capellupo14}. These effects bias
us towards more \pv\ detections in quasars that also have \siiv\ absorption.
We are also biased towards higher signal-to-noise ratio spectra. The median
signal-to-noise ratio is 6.3 and 6.6, respectively, for the `definite' and
`probable' detections, whereas it is 4.5 for the parent BAL sample.
In general, we are conservative in identifying \pv\ detections, even in the
`probable' category, and we are therefore biased towards the strongest \pv\
absorption features.  
Altogether, this means that a statistically complete sample is not possible.
While weak, moderate, and narrow \pv\ absorption lines are likely to be missed
by our search, strong \pv\ BALs are unlikely to be missed.
Instead, our goal is to identify both `definite' and `probable' detections of
\pv\ BALs for comparisons to the overall BAL population and also for
future follow-up observations of these \pv\ detections.

Overall we identify 81 `definite' and 86 `probable' \pv\ detections (out of
2694 BAL quasars), giving a firm lower limit of 3.0\%, and `probable' lower
limit of 6.2\%, for the incidence of \pv\ absorption among BAL quasars. The
SDSS coordinate names, as well as various properties from the DR9Q catalog, are
listed in Table~\ref{tab:def}, for the `definite' detections, and
Table~\ref{tab:prob}, for the `probable' detections. Rest equivalent widths
(REWs) for \civ, \siiv, and \al\ are listed in the DR9Q catalog only for
spectra with a signal-to-noise ratio of at least 5 and a BAL with
BI $>$ 500 \kms.

\section{Properties of \pv\ Quasars}
\label{sec:res}

\begin{table*}
\caption{Properties of quasars with definite \pv\ detections}
\begin{tabular}{ccccccc}
\hline
SDSS Coord. Name & $z_{em}$ & $i_{mag}$ & BI & REW(\civ) & REW(\siiv) & REW(\al) \\
  &  &  & (\kms) & (\AA) & (\AA) & (\AA) \\
\hline
J000202.16$-$004216.2	& 2.432 	& 20.0 & 3507$\pm$37 	& 25.3	& 12.0	& 0.0 \\ 
J000902.63$+$002638.5	& 2.510 	& 20.7 & 1931$\pm$28 	& 	& 	& 	 \\ 
J002709.73$+$003020.3	& 3.106 	& 19.8 & 11853$\pm$350 	& 65.0	& 39.8	& 9.4 \\ 
J004733.24$-$020315.7	& 2.890 	& 19.0 & 3686$\pm$71 	& 6.5	& 28.6	& 13.1 \\ 
J010338.44$-$020047.2	& 2.846 	& 19.3 & 7444$\pm$73 	& 41.2	& 24.4	& 0.0 \\ 
J010745.79$+$005626.3	& 2.996 	& 21.0 & 3214$\pm$139 	& 30.9	& 17.4	& 0.0 \\ 
J013652.52$+$122501.5	& 2.399 	& 17.9 & 5324$\pm$13 	& 40.8	& 13.8	& 2.5 \\ 
J013802.07$+$012424.4	& 2.530 	& 18.1 & 12463$\pm$11 	& 65.1	& 38.1	& 7.2 \\ 
J014025.63$+$002707.9	& 2.540 	& 19.6 & 8232$\pm$158 	& 55.1	& 35.5	& 7.4 \\ 
J015032.87$+$143425.6	& 4.188 	& 20.0 & 8420$\pm$271 	& 47.5	& 31.2	& 8.8 \\ 
J015221.99$+$062454.6	& 2.796 	& 19.5 & 4670$\pm$63 	& 31.9	& 23.9	& 9.0 \\ 
J015738.41$+$105705.6	& 2.807 	& 19.7 & 3523$\pm$74 	& 	& 	& 	 \\ 
J022122.51$-$044658.8	& 2.628 	& 19.8 & 2155$\pm$322 	& 	& 	& 	 \\ 
J024946.49$-$020104.1	& 2.571 	& 19.9 & 1582$\pm$48 	& 14.1	& 0.0	& 0.0 \\ 
J025000.59$-$002431.0	& 2.792 	& 20.0 & 4258$\pm$88 	& 32.2	& 15.6	& 0.0 \\ 
J025042.45$+$003536.7	& 2.397 	& 18.3 & 4851$\pm$4 	& 31.5	& 23.4	& 5.7 \\ 
J072444.07$+$392711.8	& 2.438 	& 19.6 & 2754$\pm$17 	& 8.9	& 11.8	& 4.1 \\ 
J074610.12$+$233837.5	& 3.129 	& 19.8 & 6287$\pm$859 	& 	& 	& 	 \\ 
J075145.41$+$174756.6	& 2.968 	& 20.0 & 6591$\pm$251 	& 39.3	& 28.5	& 0.0 \\ 
J075445.02$+$333215.6	& 2.690 	& 19.7 & 5444$\pm$90 	& 41.6	& 20.6	& 0.0 \\ 
J080029.38$+$124836.9	& 3.090 	& 19.4 & 5505$\pm$92 	& 41.5	& 21.6	& 0.0 \\ 
J080040.95$+$160913.0	& 3.529 	& 19.0 & 4915$\pm$268 	& 	& 	& 	 \\ 
J080814.93$+$442008.3	& 2.633 	& 20.3 & 4143$\pm$76 	& 30.4	& 20.1	& 0.0 \\ 
J081003.93$+$522507.4	& 3.895 	& 19.0 & 4651$\pm$29 	& 26.7	& 16.3	& 0.0 \\ 
J082330.71$+$111102.2	& 2.932 	& 19.6 & 3151$\pm$171 	& 33.5	& 22.8	& 6.0 \\ 
J083403.37$+$044818.3	& 3.096 	& 20.6 & 2135$\pm$245 	& 	& 	& 	 \\ 
J083509.03$+$033308.0	& 2.575 	& 20.0 & 7382$\pm$337 	& 	& 	& 	 \\ 
J084221.94$+$373331.0	& 2.381 	& 20.4 & 1190$\pm$18 	& 16.2	& 0.0	& 0.0 \\ 
J094602.54$+$364701.1	& 3.230 	& 20.2 & 4959$\pm$205 	& 	& 	& 	 \\ 
J101225.00$+$405753.3	& 3.164 	& 19.8 & 4641$\pm$206 	& 37.7	& 21.3	& 14.0 \\ 
J101324.20$+$064900.3	& 2.766 	& 18.9 & 2573$\pm$23 	& 5.0	& 17.6	& 8.5 \\ 
J102154.00$+$051646.3	& 3.439 	& 18.2 & 9456$\pm$25 	& 44.8	& 30.9	& 10.8 \\ 
J102225.90$+$041824.2	& 3.029 	& 20.5 & 1810$\pm$374 	& 	& 	& 	 \\ 
J102744.88$+$041737.5	& 2.659 	& 20.6 & 13$\pm$3 	& 	& 	& 	 \\ 
J102947.67$+$422619.5	& 2.542 	& 20.7 & 2271$\pm$127 	& 	& 	& 	 \\ 
J103220.01$+$025037.1	& 3.054 	& 20.6 & 2955$\pm$360 	& 	& 	& 	 \\ 
J103242.71$+$433605.3	& 3.491 	& 19.2 & 2685$\pm$19 	& 18.8	& 9.3	& 0.0 \\ 
J103958.20$+$061119.7	& 3.152 	& 20.2 & 848$\pm$170 	& 	& 	& 	 \\ 
J104846.63$+$440710.7	& 4.347 	& 19.6 & 12826$\pm$77 	& 65.0	& 43.8	& 9.5 \\ 
J104932.66$+$044031.4	& 2.518 	& 19.4 & 5492$\pm$56 	& 40.2	& 21.2	& 3.9 \\ 
J105233.16$-$015527.3	& 2.481 	& 19.2 & 5049$\pm$33 	& 29.5	& 15.3	& 0.0 \\ 
J105404.77$-$020931.4	& 2.705 	& 20.8 & 1236$\pm$80 	& 	& 	& 	 \\ 
J105928.52$+$011417.2	& 2.656 	& 19.2 & 3152$\pm$9 	& 23.6	& 10.7	& 0.0 \\ 
J114056.80$-$002329.9	& 3.603 	& 19.5 & 3085$\pm$86 	& 	& 	& 	 \\ 
J114740.54$+$005545.1	& 2.694 	& 19.8 & 3238$\pm$144 	& 37.7	& 14.1	& 0.0 \\ 
J114847.08$+$395544.8	& 3.001 	& 19.6 & 6970$\pm$324 	& 43.3	& 25.5	& 0.0 \\ 
J115321.80$+$371957.2	& 2.492 	& 19.3 & 7022$\pm$113 	& 43.7	& 19.3	& 0.0 \\ 
J121937.13$+$071157.7	& 2.592 	& 21.0 & 2218$\pm$135 	& 	& 	& 	 \\ 
J122124.36$+$043351.0	& 3.143 	& 18.9 & 7267$\pm$67 	& 40.6	& 26.0	& 5.9 \\ 
J124503.29$+$010929.0	& 3.219 	& 20.2 & 3213$\pm$859 	& 	& 	& 	 \\ 
J124557.16$+$344511.7	& 2.500 	& 19.6 & 2433$\pm$26 	& 0.0	& 15.1	& 6.0 \\ 
J130902.37$+$403901.5	& 2.440 	& 18.0 & 2828$\pm$19 	& 7.1	& 17.5	& 11.1 \\ 
J132604.26$+$333041.1	& 2.513 	& 19.7 & 8607$\pm$93 	& 23.7	& 27.1	& 9.6 \\ 
J135536.89$+$320323.7	& 2.772 	& 20.2 & 4300$\pm$67 	& 27.7	& 17.9	& 5.1 \\ 
J140323.37$+$040640.4	& 2.775 	& 19.9 & 1966$\pm$231 	& 	& 	& 	 \\ 
J141906.28$+$362801.9	& 2.401 	& 18.7 & 7091$\pm$15 	& 33.8	& 33.6	& 26.0 \\ 
J142102.24$+$355718.4	& 3.052 	& 20.2 & 4336$\pm$212 	& 10.0	& 15.2	& 9.5 \\ 
J142107.42$+$360920.2	& 2.663 	& 19.5 & 3749$\pm$124 	& 24.2	& 13.3	& 0.0 \\ 
J143008.04$+$063914.4	& 2.532 	& 18.4 & 2970$\pm$47 	& 13.0	& 19.1	& 5.5 \\ 
J143223.09$-$000116.4	& 2.474 	& 20.1 & 817$\pm$19 	& 11.5	& 17.1	& 0.0 \\ 
J143632.04$+$053958.9	& 2.939 	& 19.9 & 7139$\pm$204 	& 39.8	& 34.7	& 6.1 \\ 
\end{tabular}
\label{tab:def}
\end{table*}
\addtocounter{table}{-1}
\begin{table*}
\caption{continued...}
\begin{tabular}{ccccccc}
\hline
SDSS Coord. Name & $z_{em}$ & $i_{mag}$ & BI & REW(\civ) & REW(\siiv) & REW(\al) \\
  &  &  & (\kms) & (\AA) & (\AA) & (\AA) \\
\hline
J152316.17$+$335955.5	& 3.633 	& 20.0 & 3688$\pm$95 	& 	& 	& 	 \\ 
J152656.78$+$211113.8	& 2.431 	& 19.6 & 3670$\pm$38 	& 21.9	& 12.0	& 5.6 \\ 
J152657.97$+$163803.0	& 2.961 	& 20.3 & 3624$\pm$264 	& 	& 	& 	 \\ 
J152842.91$+$193105.7	& 2.668 	& 19.9 & 10863$\pm$321 	& 35.2	& 42.4	& 15.2 \\ 
J153330.03$+$222924.4	& 2.850 	& 20.2 & 5244$\pm$834 	& 	& 	& 	 \\ 
J154246.94$+$054359.3	& 2.954 	& 19.6 & 4102$\pm$143 	& 30.1	& 19.2	& 4.8 \\ 
J154727.18$+$031313.6	& 2.499 	& 19.6 & 2095$\pm$80 	& 17.0	& 14.8	& 0.0 \\ 
J154910.90$+$320620.7	& 3.078 	& 20.0 & 3625$\pm$159 	& 30.6	& 12.2	& 0.0 \\ 
J160438.61$+$181812.2	& 2.920 	& 20.2 & 6061$\pm$189 	& 38.9	& 25.7	& 5.1 \\ 
J161050.39$+$062039.5	& 2.524 	& 19.4 & 6837$\pm$46 	& 35.9	& 22.9	& 4.5 \\ 
J162554.62$+$322626.5	& 3.745 	& 19.9 & 4120$\pm$250 	& 	& 	& 	 \\ 
J164656.26$+$224454.6	& 2.532 	& 19.8 & 4002$\pm$26 	& 13.7	& 26.7	& 14.8 \\ 
J165053.78$+$250755.4	& 3.319 	& 18.6 & 4269$\pm$19 	& 35.1	& 17.9	& 7.3 \\ 
J165710.56$+$233700.2	& 2.784 	& 19.9 & 294$\pm$2 	& 	& 	& 	 \\ 
J172312.99$+$385422.7	& 2.613 	& 20.0 & 7229$\pm$197 	& 	& 	& 	 \\ 
J214855.68$-$001452.6	& 2.488 	& 19.8 & 2802$\pm$13 	& 19.7	& 3.5	& 0.0 \\ 
J230721.90$+$011118.0	& 2.780 	& 19.2 & 2892$\pm$21 	& 15.1	& 13.8	& 5.0 \\ 
J231923.84$+$004127.6	& 2.547 	& 19.1 & 3485$\pm$9 	& 21.3	& 8.5	& 0.0 \\ 
J233829.30$-$005933.8	& 3.019 	& 20.6 & 3178$\pm$302 	& 	& 	& 	 \\ 
J234058.62$+$011651.2	& 3.084 	& 20.2 & 5932$\pm$1000 	& 	& 	& 	 \\ 
\end{tabular}
\end{table*}

\begin{figure*}
 \centering
 \includegraphics[width=110mm]{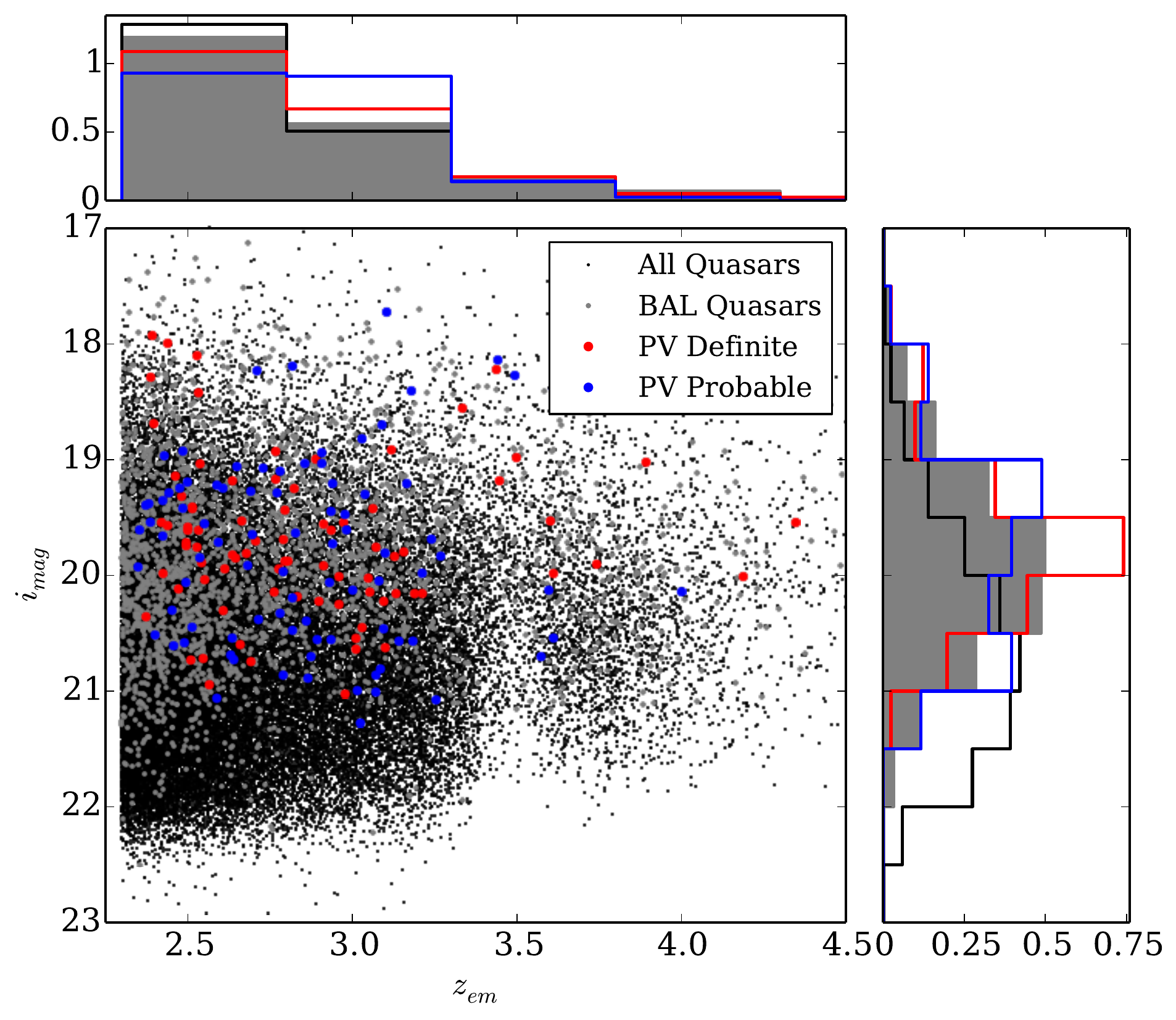}
 \caption{The distribution of $i$ magnitude with redshift, $z_{em}$. All DR9Q
   quasars with $z_{em} > 2.3$ (and S/N $>$ 0.7) are indicated as black points;
   the BAL quasars and definite and probable \pv\ detections as gray, red, and
   blue points, respectively. The top and right panels show the distribution in
   $z_{em}$ and $i$ magnitude for each sub-sample.}
 \label{fig:imag}
\end{figure*}

\subsection{Overall Sample Properties}
\label{sec:pop}

We first compare the photometric properties of the \pv-detected sample with the
overall BAL sample, using data tabulated in the DR9Q catalog.
Fig.~\ref{fig:imag} displays the distribution of $i$ magnitude (effective
wavelength of 7491\AA) with redshift. To compare with all DR9 quasars, we show
only the DR9 quasars with average signal-to-noise ratio $>$0.7 because all of
our BAL spectra have signal-to-noise above this value. The figure indicates
that the parent BAL quasar sample (gray shaded histogram) is well-matched in
redshift to the overall quasar population (black histogram). However, the
fraction of \pv\ detections (blue and red histograms) in the redshift bin at
$z$$\sim$3 is slightly higher than the parent BAL and overall quasar
populations. This result is likely a selection effect because in the lowest
redshift bin, the region of potential \pv\ absorption is near the blue end of
the spectrum, where the BOSS instrument is less sensitive. As for the
distributions in $i$ magnitude, the BAL quasars clearly tend to be brighter
than the overall quasar population. This behaviour is also likely a selection
effect because BALs are easier to detect in higher signal-to-noise spectra
\citep{Gibson09}. The distributions of `definite' and `probable' \pv\
detections, however, are consistent with the overall BAL population.

\begin{figure}
  \centering
  \includegraphics[width=72mm]{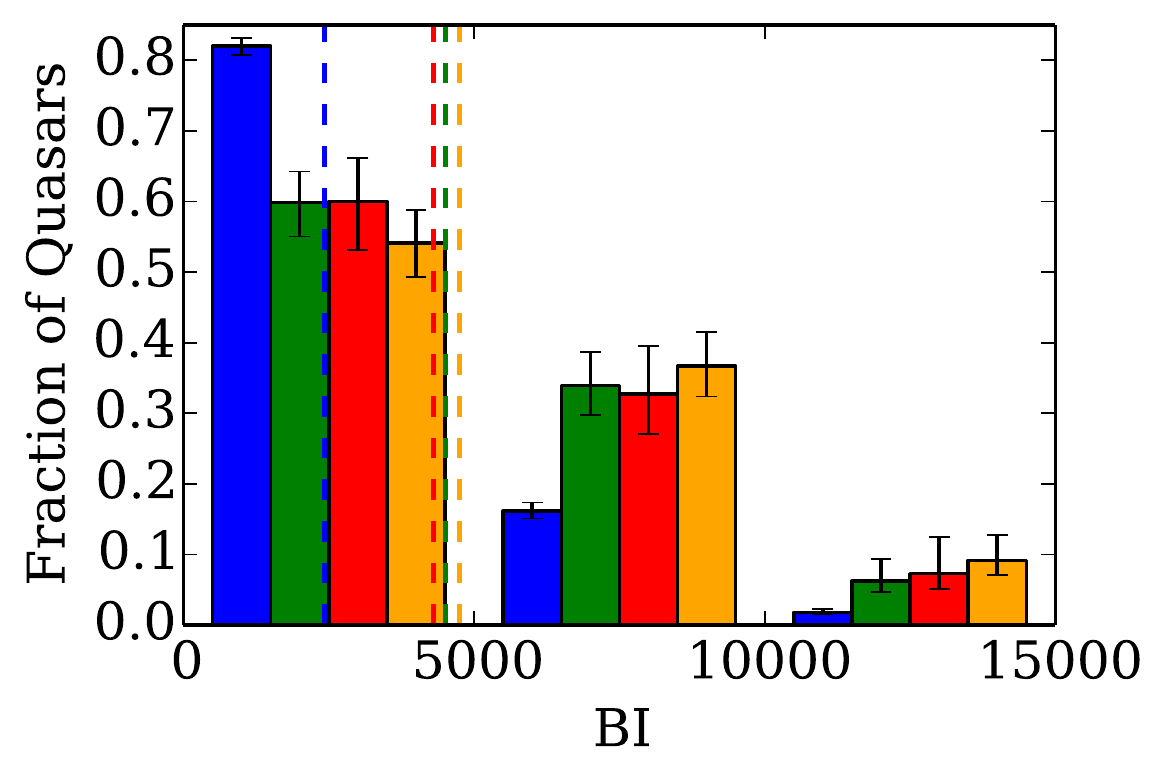}
  \includegraphics[width=72mm]{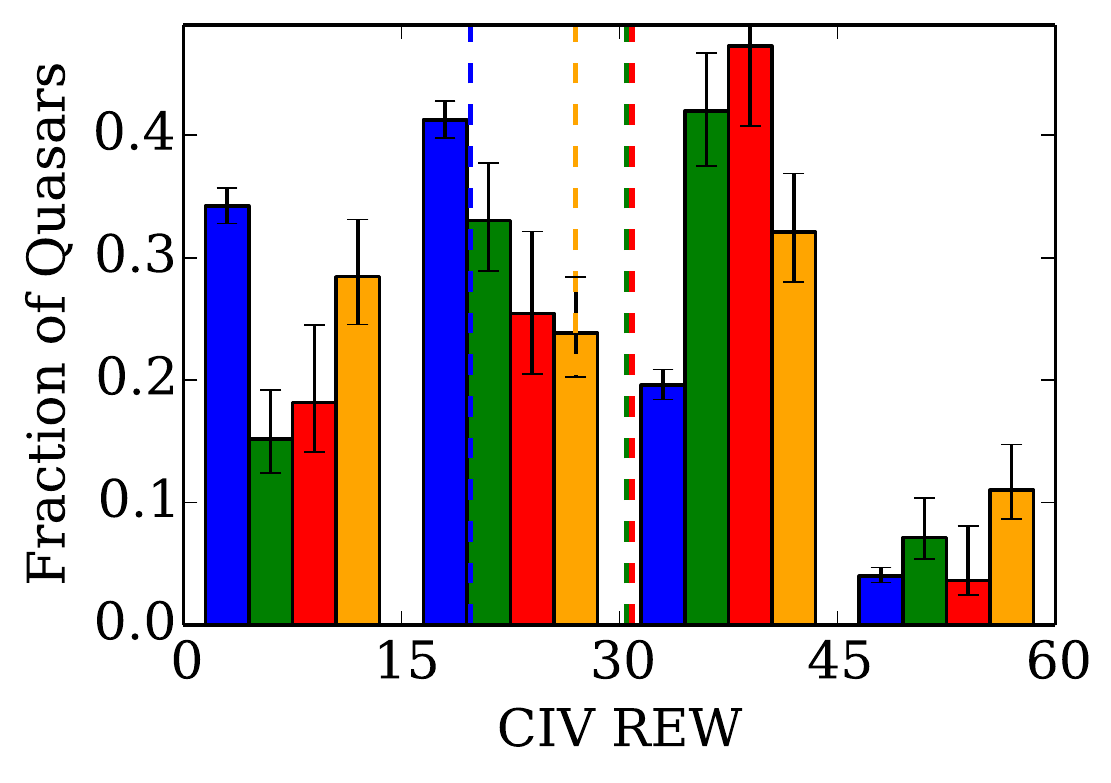}
  \includegraphics[width=72mm]{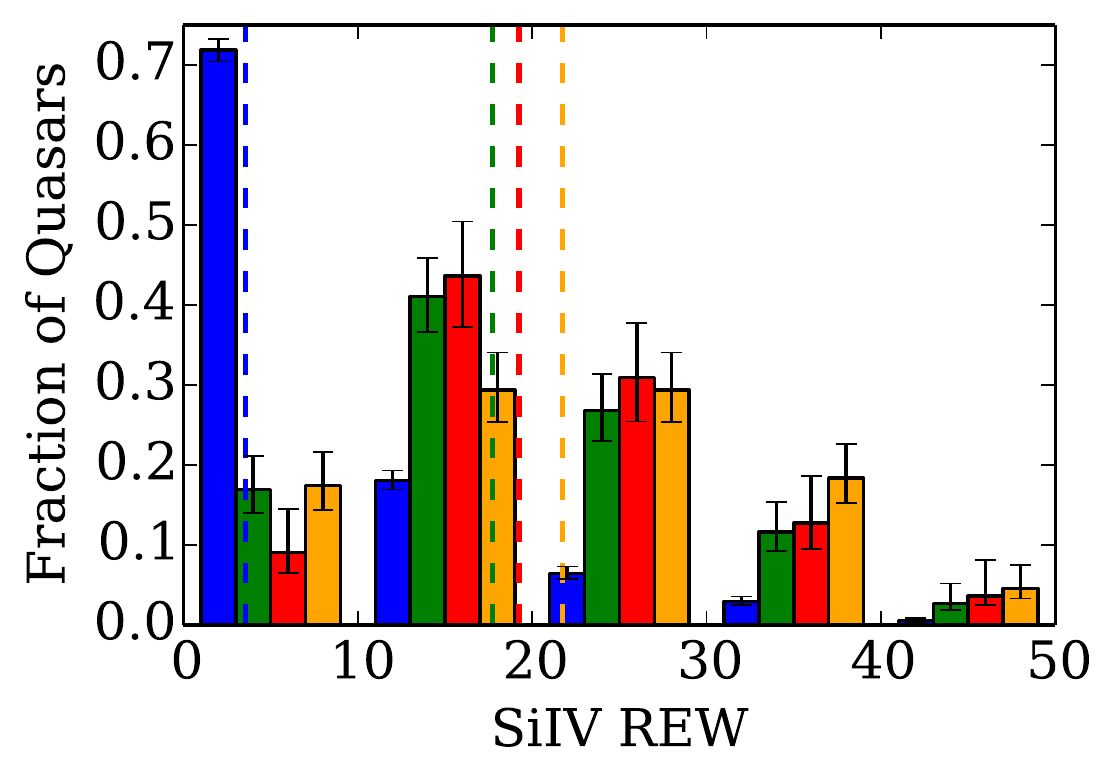}
  \includegraphics[width=72mm]{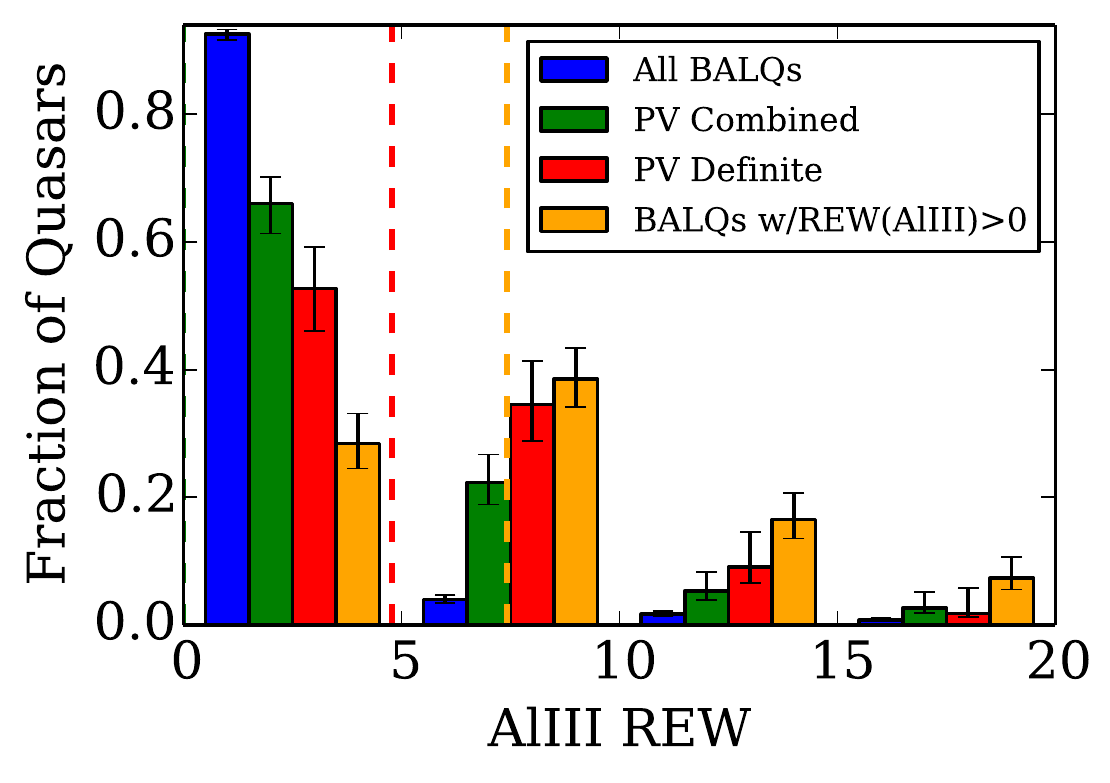}
  \caption{Histograms showing the distribution in BI (top panel) and the
    REW of \civ\ (second panel), \siiv\ (third panel), and \al\ (bottom panel)
    for the parent sample of BAL quasars (BALQs; blue) and three subsets of
    BALQs -- those with \pv\ detections (green), those with just `definite'
    \pv\ detections, and all DR9Q BALQs with a detection of \al\ absorption.
    Each set of histograms represents the fractions of sources within the full
    range
    indicated by the axis labels. This figure only includes BAL quasars with
    BI $>$ 500 \kms\ and S/N $>$ 5. The dashed lines are the median values of
    each parameter for each grouping of BALs.}
  \label{fig:hist}
\end{figure}

To compare the typical absorption line properties of the \pv-detected quasars
to the parent BAL population, we use the measurements of BI and REW tabulated
in the DR9Q catalog, listed here in Tables \ref{tab:def} and \ref{tab:prob} for
the \pv\ detections. Based on the DR9Q BI measurements, the average (median) BI
for the `definite' and `probable' detections is 4550 (4000) \kms\ and
4140 (3550) \kms, respectively. 
The average (median) BI for the parent BAL sample is 1870 (1160) \kms.

The DR9Q includes measurements of the REW of \civ, \siiv, and \al\ absorption
for those quasars with BI $>$ 500 \kms\ and signal-to-noise of at least 5 in the
rest-frame UV spectrum (with these restrictions, the median BI for the parent
BAL population is 2430 \kms\ and for the \pv\ detections is 4510 \kms).
Of the sources with `definite' and `probable' \pv\ detections with REW
measurements, 96\% have a \siiv\ REW greater than 0,
and 58\% and 32\%, respectively, have an \al\ REW greater than 0. In the overall
DR9Q BAL quasar
population, just 55\% have a \siiv\ REW greater than 0 and 10\% have an \al\
REW greater than 0 (this fraction of BOSS BAL quasars with \al\ absorption is
consistent with earlier studies that determine that LoBALs comprise $\sim$10\%
of the overall BAL population; \citealt{Trump06}). Keep in mind that we are
biased towards \pv\ detections in quasars with \siiv\ absorption because of our
visual inspection procedure (Section \ref{sec:sample}), but these results are
consistent with those of \citet{FilizAk14}, who observe many more instances of
prominent \pv\ absorption in BAL quasars with \al\ or \siiv\ BALs than those
without.

Fig. \ref{fig:hist} presents the distribution in BI, REW(\civ), REW(\siiv), and
REW(\al) for all DR9Q BAL quasars, definite \pv\ detections, combined
`definite' and `probable' detections, and BAL quasars with REW(\al) $>$ 0. The
median values of each parameter are indicated by the dashed lines, and the
1$\sigma$ error bars are based on the counting statistics for the number of
measurements in each bin \citep{Cameron11}. It is clear that the \pv\
detections are skewed towards quasars with stronger \civ\ and \siiv\ BALs, as
compared to the overall BAL population. The distributions also indicate that
the \pv\ detections tend to have similar \civ\ and \siiv\ BAL strengths as all
BAL quasars with low-ionization absorption, which we designate here as `BALQs
with REW(\al) $>$ 0.'

\begin{figure*}
 \centering
 \includegraphics[width=110mm]{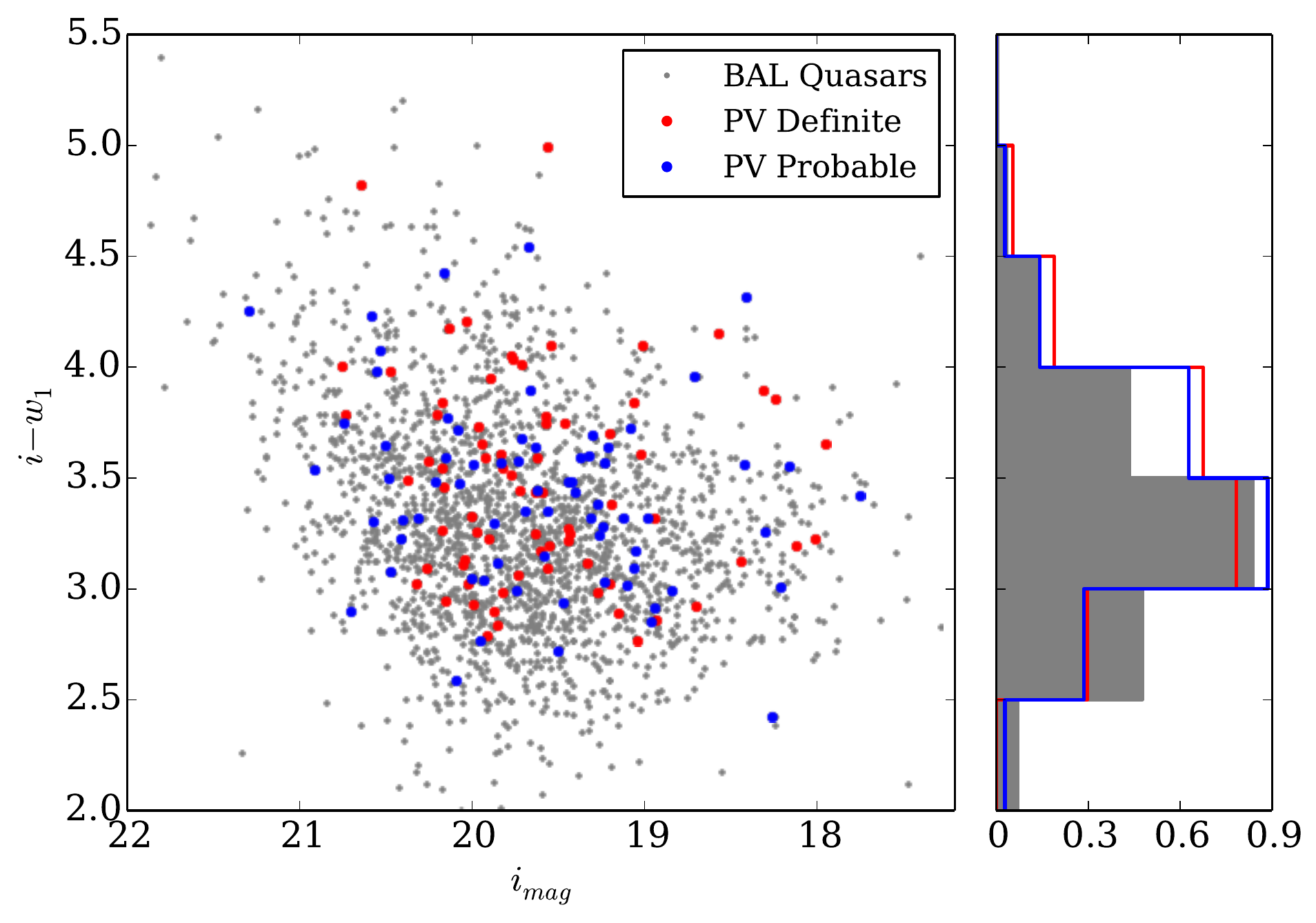}
 \caption{The same format as Fig. \ref{fig:imag}, but for $i - w1$ colour versus
   $i$ magnitude. We do not include non-BAL quasars in this figure.}
 \label{fig:imag_w1}
\end{figure*}

Fig. \ref{fig:imag_w1} displays the $i-w1$ colour versus $i$ magnitude, where
$w1$ is one of four bands of the Wide-field Infrared Survey Explorer
\citep[WISE;][]{Wright10}. The $i-w1$ colour measures the spectral slope across
the near-UV to optical wavelengths in the quasar rest-frame and is a good
indicator of dust reddening for quasars at these redshifts
(Hamann et al, in prep).
There is a slight trend towards redder colours for the \pv-detected quasars,
compared to the overall BAL quasar population. This colour
difference is probably related to the tendency for \pv-detected quasars to have
stronger absorption in low-ionization lines such as \al\ and \mgii\
(see Fig. \ref{fig:hist} and Section 4; also, Herbst et al. in preparation).
These low-ionization features are a defining characteristic of LoBALs, which
are known to have redder colours than BAL quasars without them
\citep[HiBALs;][]{Sprayberry92,Brotherton01,Trump06,Gibson09}.

We also compare the FIRST radio measurements of \pv\ detections to the overall
BAL sample. Only $3.2^{+0.1}_{-0.1} \%$ of the parent quasar population and
$3.6^{+0.4}_{-0.3} \%$ of the BAL quasars are detected by FIRST. In
comparison, $4.9^{+3.6}_{-1.5} \%$ and $3.5^{+3.2}_{-1.1} \%$ of the `definite'
and `probable' \pv-detected quasars, respectively, have FIRST detections.
Combining the `definite' and `probable' \pv-detected quasars yields
$4.2^{+2.2}_{-1.1} \%$ with FIRST detections. Given the errors due to small
number statistics for the \pv-detected quasars, the incidence of radio-loud
quasars appears to be consistent between the \pv-detected sample and the
overall (optically-selected) quasar and BAL quasar populations.

\subsection{Individual Spectra}
\label{sec:spec}

Besides examining the overall properties of the different BAL quasar
populations, it is informative to investigate the individual examples of \pv\
absorption. \pv\ absorption at the same velocity of \civ\ absorption with
nonzero intensity indicates a saturated \civ\ trough and thus that the outflow
is partially covering the AGN emission source (see Section \ref{sec:intro}).
Fig.~\ref{fig:defspec} includes three examples with a range of \pv\ profile
shapes, from broad and deep like the corresponding \civ\ BAL to a clearly
identifiable doublet. In all three cases, the \civ\ feature reaches zero, or
nearly zero, intensity.
Fig.~\ref{fig:defspec2} shows three examples where the \civ\ BAL clearly does
not reach zero intensity.  The average covering fractions
for the \civ\ absorption in these 3
cases ranges from $\sim$0.4 to 0.6. In quasars such as J103242+433605, \pv\
absorption across a velocity range where \civ\ has a rounded or complex profile
implies that this \civ\ profile shape is due to a velocity-dependent covering
fraction, and not velocity-dependent optical depths.

Despite our bias towards finding only strong \pv\ BALs (see
Section~\ref{sec:sample}), we observe that, in general, the \pv\ absorption is
not quite as deep as the corresponding \civ\ and \siiv\ absorption (see, for
example, J013802+012424 in Fig.~\ref{fig:defspec} and J151211+012807 in
Fig.~\ref{fig:probspec}). However, the \pv\ absorption is generally similar in
width to the corresponding \civ\ and \siiv\ BALs, as seen in all
the spectra in Figs.~\ref{fig:defspec}~to~\ref{fig:probspec}, at least in
cases where it is not a resolved doublet.
The depths of the \civ\ and \pv\ BALs in the examples in
Fig.~\ref{fig:defspec2} are similar. This result implies that the \pv\ trough
is also saturated with a covering fraction similar to \civ.
There is some uncertainty in the depth of the \pv\ absorption due to the
possibility of unrelated Ly$\alpha$ forest lines affecting the profile.
However, in cases such as J124557+344511, where the shape of the \pv\
absorption profile closely matches the \civ\ and/or \siiv\ profile, it is clear
that the profile is shaped by a velocity-dependent covering fraction and not
intervening absorption.

In general, the \pv\ absorption profiles more closely resemble \siiv\
than \civ. For example, in J214855$-$001452, both \siiv\ and \pv\ have a
clearly visible doublet, while the \civ\ spans a wider range in velocities.
There are also cases such as J013802+012424 where the \civ\ BAL has a strong
wing extending blueward, but this wing is less pronounced in \siiv\ and not
seen in \pv. However, there is also J081410+323225 (Fig.~\ref{fig:probspec}),
where the flux within the \pv\ absorption feature goes to zero, as in the \civ\
trough, but the \siiv\ absorption is much weaker.

Finally, the \pv\ absorption in J013802+012424 (Fig.~\ref{fig:defspec})
is a doublet that appears to have the same depth in both components, indicating
saturated absorption and optical depth $>$$\sim$3. This puts the optical depth
of \civ\ at $>$1000 and gives very high values of the total column density in
the outflow (see Section \ref{sec:discuss} for further discussion).

\section{Composite Spectra}
\label{sec:comp}

\begin{figure}
  \centering
  \includegraphics[width=80mm]{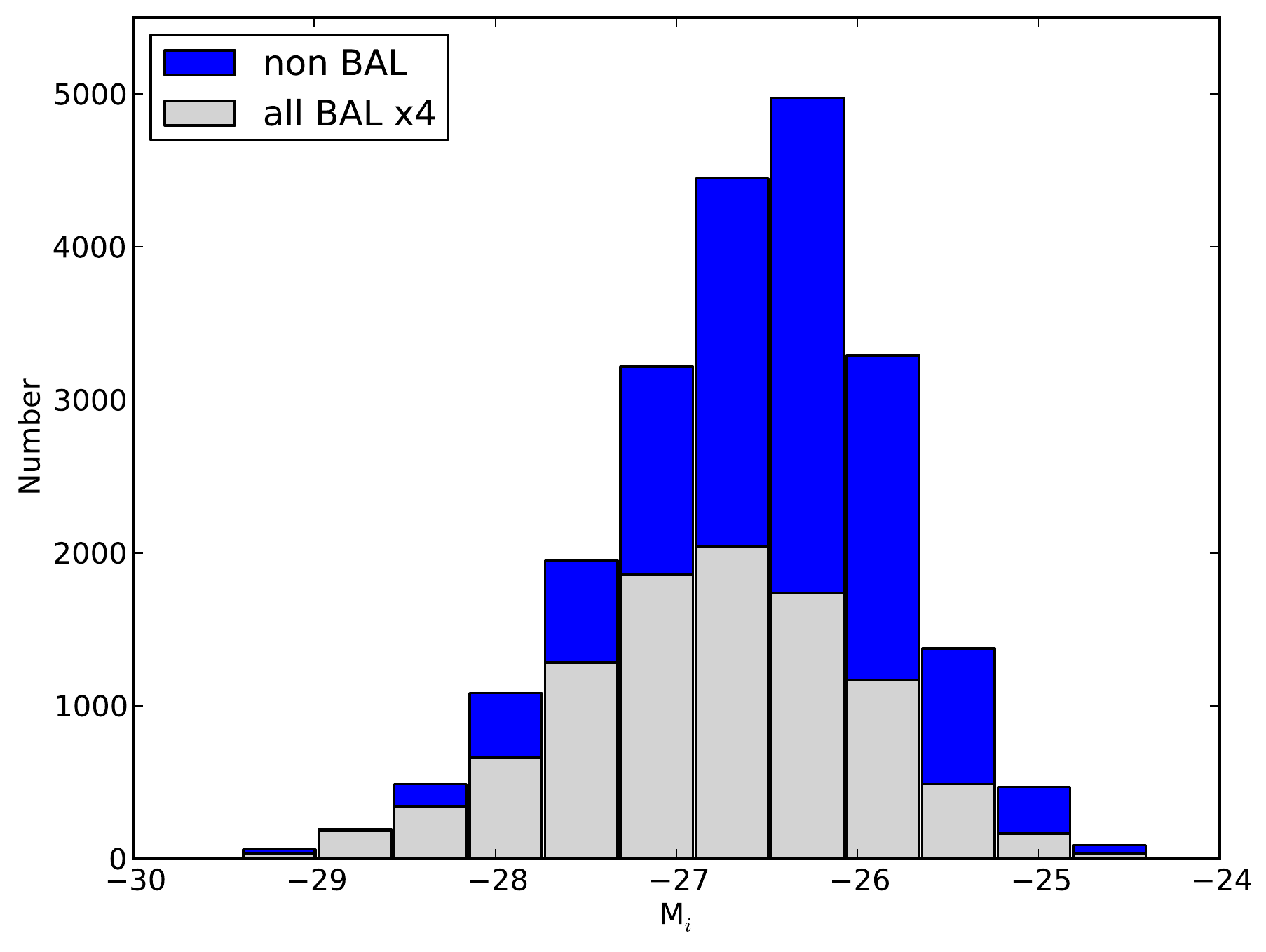}
  \caption{Comparison of the absolute magnitude of the non-BAL and BAL quasar
    populations used for generating the composite spectra.}
  \label{fig:Mi}
\end{figure}

\begin{figure*}
  \centering
  \includegraphics[width=120mm]{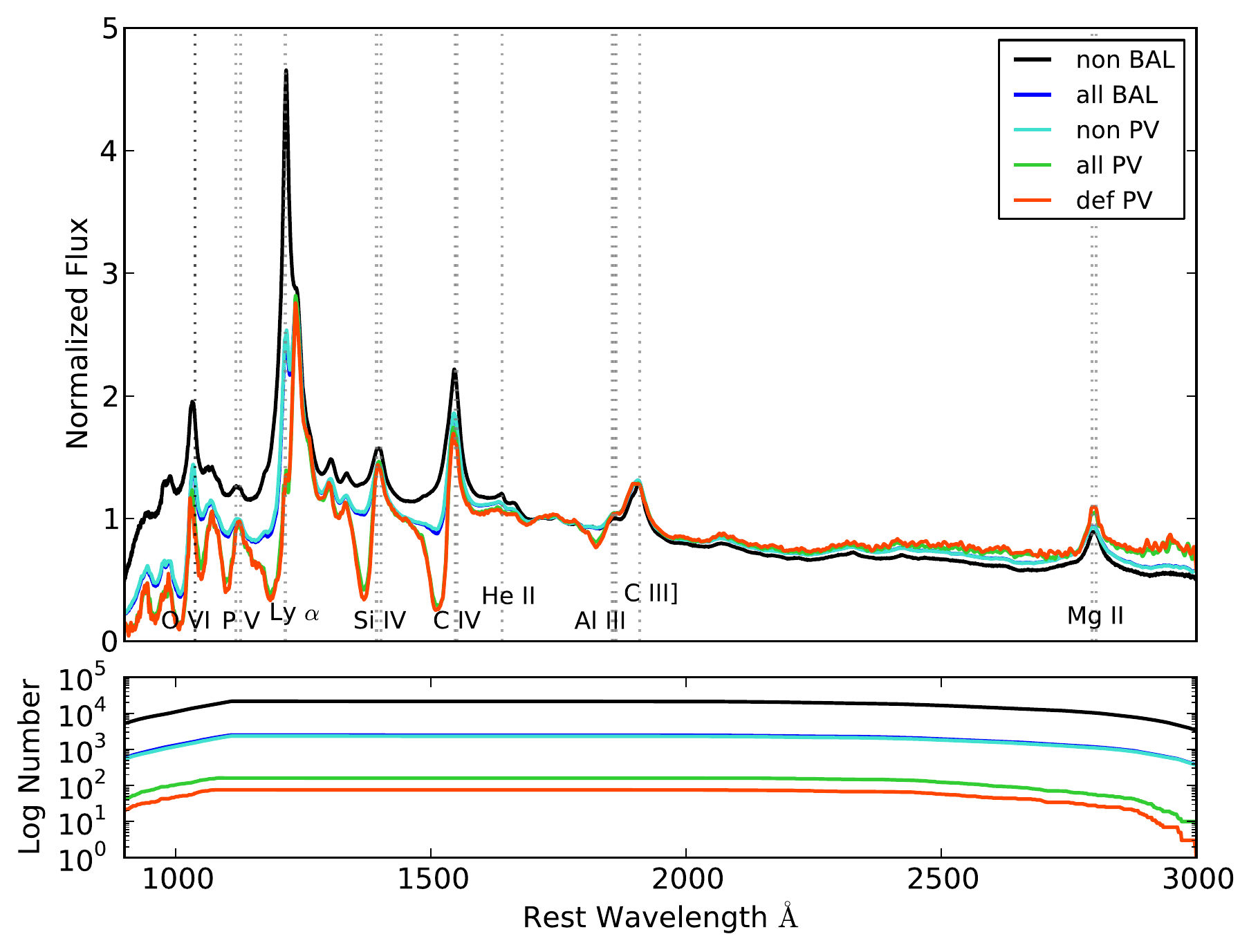}
  \caption{The top panel displays the composite spectra for non-BAL quasars
    (black curve), BAL quasars (blue curve), BAL quasars without a \pv\
    detection (light blue curve), BAL quasars with either a `definite' or
    `probable' \pv\ detection (green curve), and BAL quasars with a `definite'
    \pv\ detection (red curve). Several prominent emission lines are marked by
    dotted lines and labeled across the bottom. The bottom panel shows the
    number of quasar spectra included in each composite at each wavelength.}
  \label{fig:comp_full}
\end{figure*}

\begin{figure}
  \centering
  \includegraphics[width=85mm]{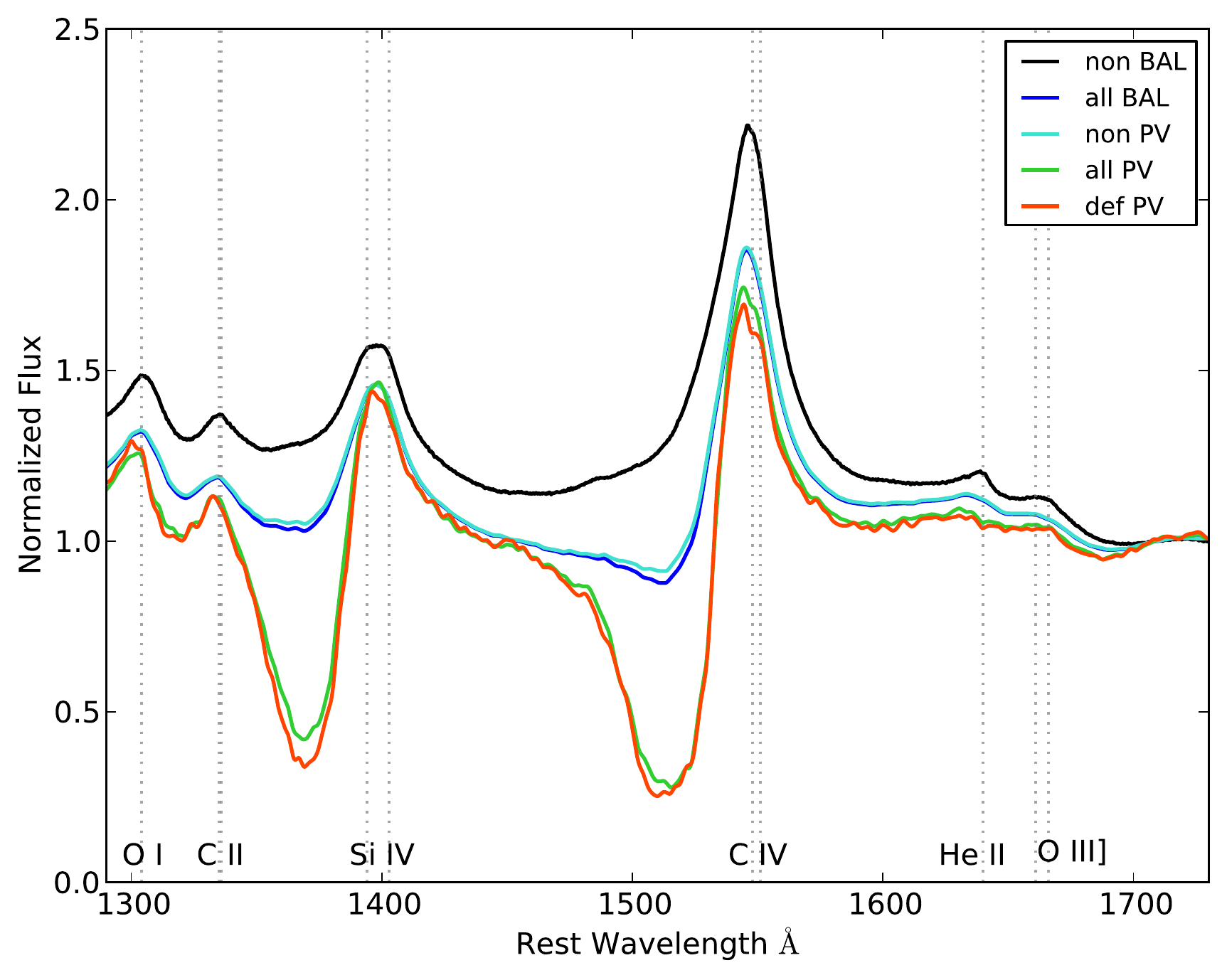}
  \caption{The \siiv\ and \civ\ region of the composite spectra shown in
    Fig.~\ref{fig:comp_full}.}
  \label{fig:comp_siiv_civ}
\end{figure}

\begin{figure*}
  \centering
  \includegraphics[width=120mm]{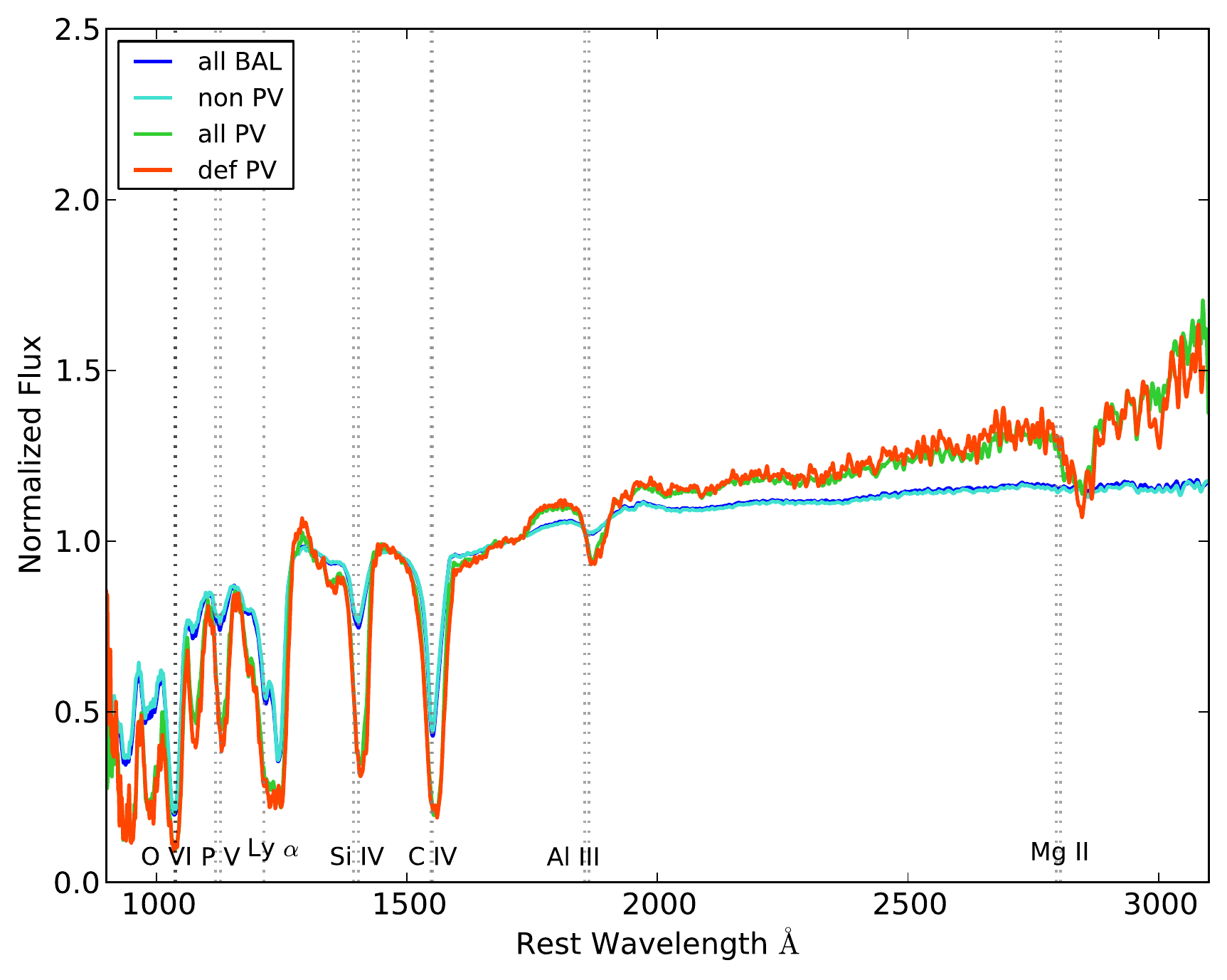}
  \caption{Modified versions of the BAL composite spectra shown in
    Fig.~\ref{fig:comp_full}. Here, the individual BAL spectra are
    shifted to the absorber frame, and then the resulting composites are
    divided by modified non-BAL composites with matching distributions of
    velocity shifts.}
  \label{fig:comp_full_norm}
\end{figure*}

\begin{figure}
  \centering
  \includegraphics[width=85mm]{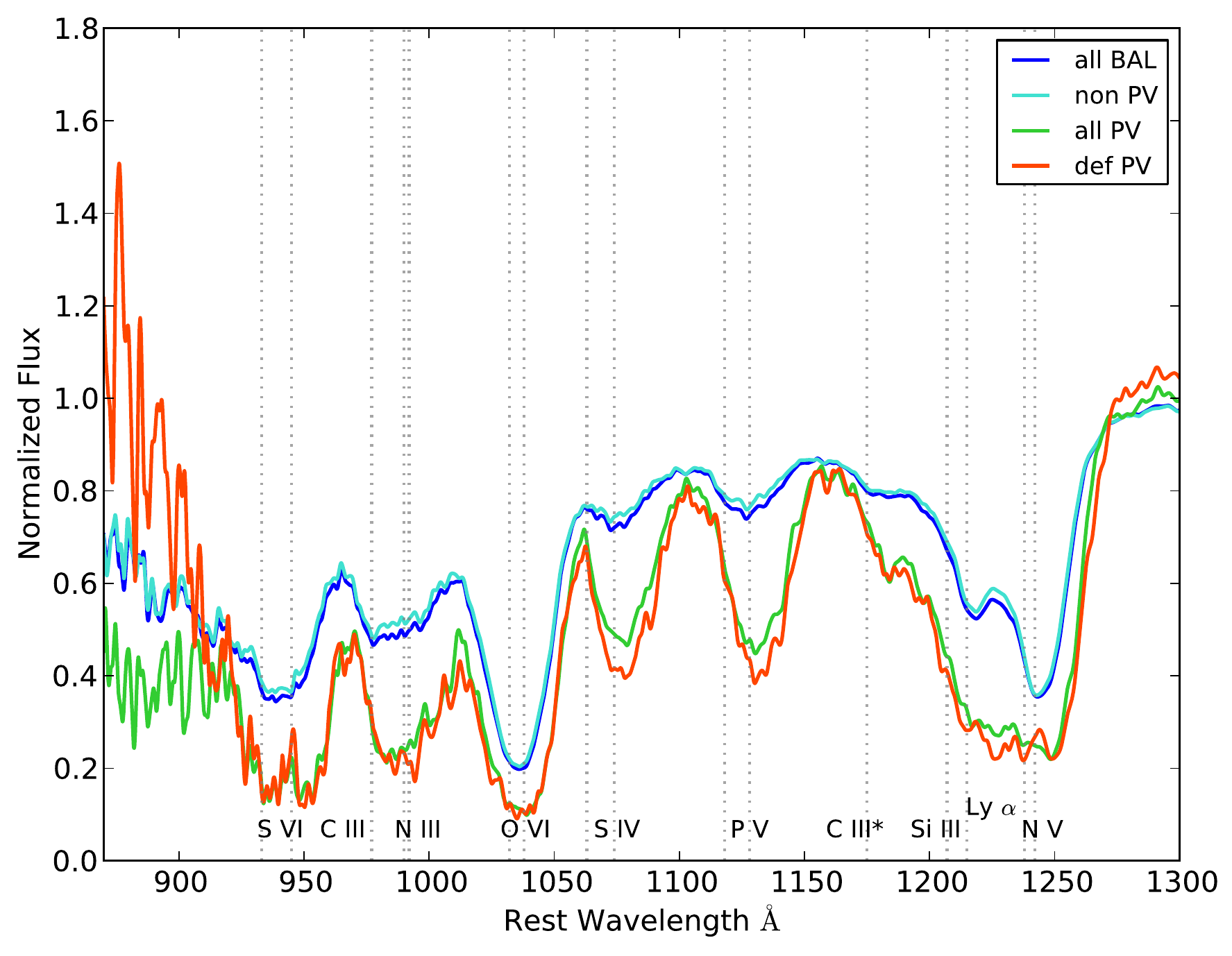}
  \includegraphics[width=85mm]{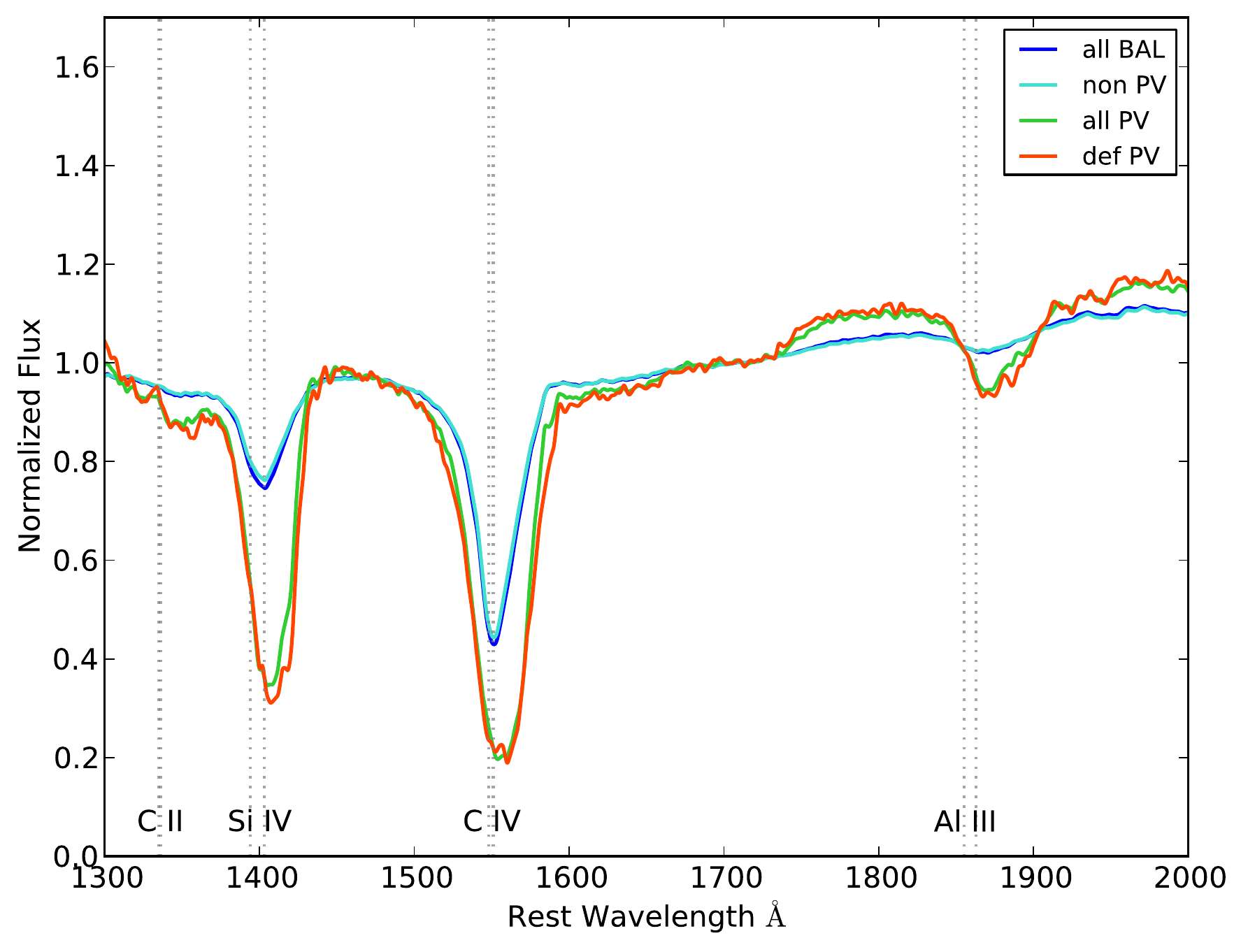}
  \caption{An expanded view of the region blueward of Ly$\alpha$ (top) and the
    region containing \siiv, \civ, and \al\ (bottom) in the composite spectra
    shown in Fig.~\ref{fig:comp_full_norm}.}
  \label{fig:comp_norm}
\end{figure}

\begin{figure}
  \centering
  \includegraphics[width=85mm]{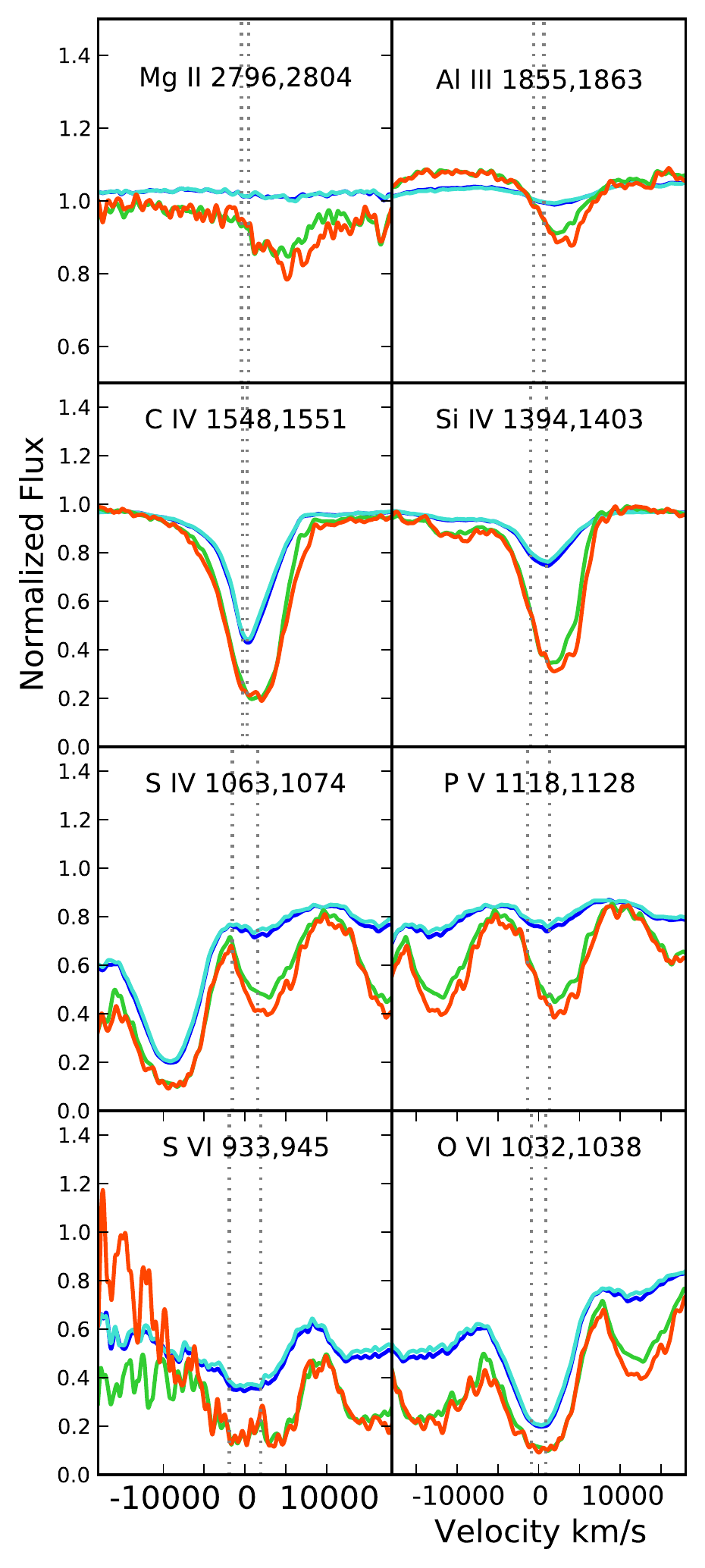}
  \caption{A grid of individual absorption features in the composite spectra
    shown in Figs~\ref{fig:comp_full_norm} and \ref{fig:comp_norm}.
    Zero velocity is set to the midpoint in each of these doublets. The dotted
    vertical lines are the nominal positions of the individual doublet
    components.
    }
  \label{fig:comp_balgrid}
\end{figure}

An approach to compare the overall properties of BAL quasars with \pv\
detections to those without is to create composite spectra of each group. One
strong advantage to analyzing composite spectra is that Ly$\alpha$ forest lines
in individual quasars average out in the composites, and, therefore, it becomes
easier to identify intrinsic absorption of the different species blueward of
Ly$\alpha$. Several earlier works, such as \citet{Reichard03a},
\citet{Baskin13}, and \citet{FilizAk14}, have created composites of BAL
quasars, and we expand on these earlier works with composites of quasars with
clear \pv\ absorption. We follow the procedures described in Herbst et al. in
prep for generating the composites.

There are known flux calibration issues with the BOSS spectra
\citep[e.g.][]{Paris11}, so prior to creating the composite spectra, we apply
a correction based on airmass to the individual spectra. The flux calibration
issues and flux correction are described in \citet{Harris16}. While the
corrections for individual quasars are uncertain, they are expected to be
accurate, on average, for samples of quasars that are randomly located on the
sky (e.g., in our composites).

We first create a composite of non-BAL quasars. To match properly the sample of
BAL quasars with a sample of non-BAL quasars, we remove some of the fainter
quasars to best match the distributions in absolute $i$-magnitude ($M_i$). This
is intended to remove any luminosity-related biases, e.g. the Baldwin effect
\citep{Baldwin77}. We first impose a cut in signal-to-noise ratio. We
experiment with different values of signal-to-noise ratio for this cutoff, and
we find that requiring a signal-to-noise ratio of at least 1.5 produces the
closest match between the $M_i$ distributions of BALs and non-BALs, without
removing any of the definite detections of \pv.
This signal-to-noise ratio cut reduces the difference in median $M_i$ between
the BAL and non-BAL samples from 0.87 to 0.50 mag. To improve further the match
in $M_i$, we randomly remove non-BAL quasars from the fainter bins until we
obtain the best match between the two populations. This further reduces the
difference in median $M_i$ between the two samples to 0.24 mag.
Fig.~\ref{fig:Mi} presents the resulting distributions of the non-BAL (blue)
and BAL (grey) populations. With the exception of the additional
signal-to-noise ratio cut, this is the same BAL population as described in
Section~\ref{sec:sample}.

We next normalize all the individual spectra by the continuum flux at
$\sim$1700\AA\ and take the median of all the spectra in these two
populations. We compare these composites to the
composites of three subgroups of BAL quasar spectra:
the spectra of BAL quasars without detected \pv\ absorption, all BAL quasars
where we detect \pv\ absorption, and only the definite detections of \pv\
absorption. The results are displayed in Fig. \ref{fig:comp_full}.

An expansion of the \siiv\ and \civ\ absorption region is shown in
Fig.~\ref{fig:comp_siiv_civ}. It is immediately clear that our \pv\ detections
tend to occur in those quasars with strong \civ\ and \siiv\ BALs, as indicated
already by the examples in Figs \ref{fig:defspec} to \ref{fig:probspec} and
also in \citet{FilizAk14}. The \civ\ and \siiv\ absorption in the parent BAL
population and in the \pv-detected quasars appear to occur at similar
velocities and have similar shapes. The BALs are simply deeper in the
\pv-detected quasars. However, as discussed in Section \ref{sec:sample},
we are biased towards detecting only the strongest \pv\ absorption features, as
we are searching in the Ly$\alpha$ forest.

It is also clear that non-BAL quasars tend to have stronger broad emission
lines, including notably \civ\ and \heii\ $\lambda$1640. In fact, there appears
to be a progression from strong emission lines in non-BAL quasars, to moderate
lines in the non-\pv-detected BAL quasars, to the weakest lines in the
\pv-detected BAL quasars (again, we try to closely match the distribution in
$M_i$ between the non-BAL and BAL samples in order to avoid biases like the
Baldwin effect). This behaviour, particularly in \heii, is probably tied
to a trend in the far-UV spectral energy distributions that increasingly favors
radiative acceleration in higher column density BAL outflows with \pv\ (see
Section \ref{sec:discuss} below).

To analyze further the composite absorption troughs, we use the non-BAL spectra
to remove the emission features. We first shift the BAL spectra to align the
centers of the \civ\ BAL troughs, using the midpoint between the minimum
(VMIN\_CIV\_2000) and maximum (VMAX\_CIV\_2000) velocity values of the troughs
given in the DR9Q catalog \citep{Paris12}, to create absorber-frame
composites. The non-BAL spectra are then shifted by the same distribution of
shifts as the BAL spectra for each BAL category. This shifting blurs the
emission features in the same way as the corresponding absorber-frame BAL
composites, and dividing by the appropriately shifted non-BAL composite removes
the emission features. The results are presented in Fig. \ref{fig:comp_full_norm},
with the same colour scheme as Fig. \ref{fig:comp_full}.

Fig.~\ref{fig:comp_full_norm} demonstrates that the BAL quasars with \pv\
detections tend to have redder spectra than the overall BAL population, in
agreement with Fig.~\ref{fig:imag_w1}, which shows that \pv-detected quasars
tend to be skewed toward redder colours as compared to the overall BAL
population.

Fig. \ref{fig:comp_full_norm} also indicates a propensity for \pv\ absorption
to appear in LoBAL quasars. This is indicated by the clear \al\ and \mgii\
absorption features in the \pv\ quasar composites. In the overall BAL
composite, the dip at \al\ is much shallower than in the \pv\ composites, and
there is no visible dip corresponding to \mgii\ absorption. This is all in
agreement with the results of Section~\ref{sec:res}, where it is shown that the
incidence of \al\ absorption is much higher among the \pv-detected quasars than
in the overall BAL quasar population.

Finally, Fig. \ref{fig:comp_norm} presents expanded views of two regions of the
composite spectrum of Fig. \ref{fig:comp_full_norm}; the top panel is the
region blueward of 1300\AA, and the bottom panel is the region redward of
1300\AA. Fig. \ref{fig:comp_balgrid} plots a selection of individual BALs from
the normalized composite BAL spectra. In general, all of the lines
(\mgii, \al, \civ, \siiv, \siv, \pv, \svi) are much deeper in the \pv\
detections than in the parent BAL population. While the depth of \ovi\ is also
larger in the \pv\ composites, the difference is not as striking as for the
other lines. The bottom of the \civ\ and \siiv\ lines are also skewed toward
different velocities in the all-BAL versus \pv-only composites
(Fig.~\ref{fig:comp_balgrid}). The same effect may be present in \al\ and \pv,
but the absorption in the all-BAL composites is too weak to make a firm
conclusion. The \ovi\ line, on the other hand, appears to be centered at the
same velocities in all of the composites.

The \pv\ absorption in the `definite' composite is only slightly stronger than
in the total \pv\ composite, and the shapes of the \pv\ feature are quite
similar. This likely indicates that most, if not all, of the `probable'
detections are indeed actual \pv\ detections. Furthermore, there is a clear
depression in both the total BAL composite (dark blue curve) and the composite
of BAL quasars without a \pv\ detection (light blue curve) in the \pv\ region.
The fact that excluding our list of \pv\ detections from the total BAL
composite does not remove this depression indicates that our list of \pv\
detections is missing, perhaps many, quasars with weaker \pv\ features.

\section{Discussion}
\label{sec:discuss}

We have identified 81 `definite' and 86 `probable' detections of \pv\
absorption, from a parent sample of 2694 BAL quasars from the SDSS-III BOSS
DR9 quasar catalog. While the \pv-detected quasars are distributed similarly in
redshift and luminosity as the overall BAL population (Fig. \ref{fig:imag}),
they tend to have slightly redder colours (Fig. \ref{fig:imag_w1}).

While our selection procedure and contamination by the Ly$\alpha$ forest
produces a bias towards finding only strong \pv\ absorption, the
absorption in other species, such as \civ\ and \siiv, is generally stronger
than in the average BAL quasar. This is seen clearly when comparing the BI and
REW distribution of all BAL quasars to just the quasars with \pv\ detections
(Fig. \ref{fig:hist}), as well as in the composite spectra of the two
populations (e.g. Fig. \ref{fig:comp_siiv_civ}). In fact, \siiv\ is present in
nearly all (96\%) of our \pv\ detections, whereas it is only detected in 55\%
of the parent BAL population (Section \ref{sec:pop}; see also \citealt{FilizAk14}).

\pv, and to a lesser extent \siv\ and \svi, are low-abundance ions, and
detections of these lines are important for constraining outflow energetics.
The presence of absorption in these low abundance ions in BAL spectra indicates
that other lines, such as \civ\ and \siiv\ are saturated.
Previous studies using photoionization modeling, and the assumption of
solar abundances, have found that the optical depth in \civ\ is much higher
than in these low abundance ions. In particular, \citet{Dunn12} show that the
optical depth of \civ\ is $\sim$30 to 40 times greater than the optical depth
of \siv, and various works have found that the \civ\ optical depth is at least
$\sim$100 to $\sim$1200 times greater than the optical depth of \pv\
\citep{Hamann98,Leighly09,Leighly11,Borguet12}.
Thus, the presence of \pv\ and \siv\ indicates saturated \civ\ lines, and
saturated absorption lines are shaped by the covering fraction of the outflow
relative to the continuum radiation source, and not by the optical depth of the
line. Therefore, in cases where \civ\ does not reach zero intensity at the same
outflow velocities as the \pv\ and/or \siv\ absorption, the outflow is only
partially covering the continuum source. All of this indicates that the column
densities of the flows are much larger than inferred based on the \civ\ profile
alone, with values of log $N_H$ determined to be $>$22 \cmN.

It is possible, however, that absorption lines that do not reach zero intensity
do completely cover the background continuum source, but the BAL troughs do not
reach zero because of scattered continuum light from a more extended region.
This idea is consistent with some polarization data that reveal, in some cases,
higher degrees of polarization in BAL troughs compared to the adjacent
continuum \citep[][and refs therein]{Ogle99,Lamy00,Lamy04,Brotherton06}.
However, we present examples where \pv\ absorption indicates saturated \civ\
absorption and thus a velocity-dependent covering fraction across the profile.
This is inconsistent with the scenario of scattered light filling in the BAL
troughs because it is unlikely that scattered light would produce such a
velocity-dependent structure. In any case, whether or not the outflow
completely covers the background source, the \pv\ absorption still indicates
saturated \civ\ absorption and large column densities.

\citet{Capellupo14} use an estimate of the column density in one quasar, based
on the detection of strong \pv\ absorption, along with an estimate of the
distance of the flow based on corresponding variability in \pv\ and clearly
saturated \civ\ absorption, to estimate the energetics of the flow. A column
density, $N_H$, of $2 \times 10^{22}$ \cmN\ and a distance of 3.5 pc give a
total mass of 4100 $M_{\sun}$, an outflow rate of 12 M$_{\sun}$ yr$^{-1}$, and
a kinetic energy luminosity of $4 \times 10^{44}$ ergs s$^{-1}$. This value is
roughly 2\% of the bolometric luminosity of the quasar, and is similar to the
values quoted for a flow to have a significant impact on its host galaxy via
feedback \citep[e.g.][]{Scannapieco04,Hopkins10}.
The \pv\ absorption in \citet{Capellupo14} has an estimated optical
depth up to $\sim$1.5. In the current work, where we identify cases in which
the \pv\ absorption has a resolved doublet with a one-to-one depth ratio (e.g.,
J013802+012424), the optical depth can be at least 2 times larger, giving even
larger column densities, mass-loss rates, and kinetic energy than found in
\citet{Capellupo14}.

In the current work, we find `definite' or `probable' detections of \pv\ in
$\sim$6\% of the BAL quasars in our parent sample. Furthermore, the presence of
absorption at the location of \pv\ in the composite spectrum of all BAL quasars
minus these \pv\ detections (Fig. \ref{fig:comp_balgrid}, light blue curve)
indicates there are a number of instances of \pv\ absorption that we missed.
In fact, Herbst et al., in preparation, detect \pv\ absorption even in
composite spectra of just weak \civ\ BALs, indicating that \pv\ absorption may
be more common than our individual \pv\ detections indicate.
It is also unknown whether the detection of \pv\ absorption depends on
our line of sight to the flow. Therefore, a significant number of BAL outflows
might have column densities similar to those of the quasar studied in
\citet{Capellupo14}, whether or not they show \pv\ absorption
(although, BAL variability studies often show BALs at different velocities 
varying in concert, which supports the idea of a lower column density outflow
that is more susceptible to changes in the global ionizing flux; see e.g.
\citealt{FilizAk14}).

With the detection of \pv\ giving a lower limit on the column density of the
flow, the main remaining variable in determining the energetics is the distance
of the flow. Outflows likely exist at a range of distances, and larger
distances yield higher kinetic luminosity estimates for a given column density
\citep[e.g.][]{Capellupo14,Chamberlain15}.

As mentioned in Section \ref{sec:comp}, the composite spectra show a weakening
\heii\ emission line from the non-BAL quasar population to the BAL quasars to
the \pv\ quasars. A weaker \heii\ suggests there are fewer far-UV photons
available to produce the \heii\ line, indicating a softer far-UV SED. Moreover,
there is evidence of intrinsic X-ray weakness in some BAL quasars, as compared
to non-BAL quasars \citep{Luo14}. A harder spectrum would more highly ionize
the outflowing gas, lowering its opacities and making it more difficult to
accelerate \citep{Baskin13}. Our results suggest that a softer far-UV
continuum, identified by weaker \heii\ emission, can help to drive larger
column density, more powerful outflows.

The skew towards redder colours in \pv-detected quasars, as well as a much
higher incidence of low-ionization absorption among \pv-detected quasars,
indicates an overlap between the populations of LoBAL quasars and quasars with
strong \pv\ absorption, as LoBALs also tend to have redder colours than the
overall BAL population \citep{Gibson09}. LoBALs are sometimes attributed to a
younger stage of quasar evolution, with higher accretion rates and, perhaps,
more powerful outflows in young, dusty host galaxies
\citep[][and references therein]{Urrutia09,Glikman12}. Another possibility is
that LoBALs are the result of orientation effects, where lower degrees of
ionization and more dust extinction appear along lines of sight nearer the
accretion disc plane (\citealt{Hamann04,Baskin13}; and figure 16 of
\citealt{FilizAk14}). Either of these interpretations could apply to quasars
with \pv\ absorption. The large column densities and more powerful outflows
that \pv\ absorption suggests could occur either in younger quasars or along
sightlines closer to the accretion disc plane.

Further investigations utilizing the large sample of \pv\ detections in this
work will help to answer some of the remaining questions on the typical
energetics of BAL flows.

\section*{Acknowledgments}

We thank the referee for helpful comments on the manuscript.
FH and HH acknowledge support from the USA National Science Foundation (NSF)
grant AST-1009628.
WNB acknowledges support from NSF grant AST-1516784.

Funding for SDSS-III has been provided by the Alfred P. Sloan Foundation, the Participating Institutions, the National Science Foundation, and the U.S. Department of Energy Office of Science. The SDSS-III web site is http://www.sdss3.org/.

SDSS-III is managed by the Astrophysical Research Consortium for the Participating Institutions of the SDSS-III Collaboration including the University of Arizona, the Brazilian Participation Group, Brookhaven National Laboratory, Carnegie Mellon University, University of Florida, the French Participation Group, the German Participation Group, Harvard University, the Instituto de Astrofisica de Canarias, the Michigan State/Notre Dame/JINA Participation Group, Johns Hopkins University, Lawrence Berkeley National Laboratory, Max Planck Institute for Astrophysics, Max Planck Institute for Extraterrestrial Physics, New Mexico State University, New York University, Ohio State University, Pennsylvania State University, University of Portsmouth, Princeton University, the Spanish Participation Group, University of Tokyo, University of Utah, Vanderbilt University, University of Virginia, University of Washington, and Yale University.

\bibliographystyle{mn2e}

\bibliography{full_bibliography_bals}

\begin{thebibliography}{}

\bibitem[\protect\citeauthoryear{{Ahn}, {Alexandroff}, {Allende Prieto},
  {Anderson}, {Anderton}, {Andrews}, {Aubourg}, {Bailey}, {Balbinot}, {Barnes}
  \& et al.}{{Ahn} et~al.}{2012}]{Ahn12}
{Ahn} C.~P.,  {Alexandroff} R.,  {Allende Prieto} C.,  {Anderson} S.~F.,
  {Anderton} T.,  {Andrews} B.~H.,  {Aubourg} {\'E}.,  {Bailey} S.,  {Balbinot}
  E.,  {Barnes} R.,    et al. 2012, \apjs, 203, 21

\bibitem[\protect\citeauthoryear{{Arav}, {Becker}, {Laurent-Muehleisen},
  {Gregg}, {White}, {Brotherton} \& {de Kool}}{{Arav} et~al.}{1999}]{Arav99a}
{Arav} N.,  {Becker} R.~H.,  {Laurent-Muehleisen} S.~A.,  {Gregg} M.~D.,
  {White} R.~L.,  {Brotherton} M.~S.,    {de Kool} M.,  1999, ApJ, 524, 566

\bibitem[\protect\citeauthoryear{{Asplund}, {Grevesse}, {Sauval} \&
  {Scott}}{{Asplund} et~al.}{2009}]{Asplund09}
{Asplund} M.,  {Grevesse} N.,  {Sauval} A.~J.,    {Scott} P.,  2009, ARA\&A,
  47, 481

\bibitem[\protect\citeauthoryear{{Baldwin}}{{Baldwin}}{1977}]{Baldwin77}
{Baldwin} J.~A.,  1977, \apj, 214, 679

\bibitem[\protect\citeauthoryear{{Baskin}, {Laor} \& {Hamann}}{{Baskin}
  et~al.}{2013}]{Baskin13}
{Baskin} A.,  {Laor} A.,    {Hamann} F.,  2013, \mnras, 432, 1525

\bibitem[\protect\citeauthoryear{{Bolton}, {Schlegel}, {Aubourg}, {Bailey},
  {Bhardwaj}, {Brownstein}, {Burles}, {Chen}, {Dawson}, {Eisenstein}, {Gunn},
  {Knapp} \& {et al.}}{{Bolton} et~al.}{2012}]{Bolton12}
{Bolton} A.~S.,  {Schlegel} D.~J.,  {Aubourg} {\'E}.,  {Bailey} S.,  {Bhardwaj}
  V.,  {Brownstein} J.~R.,  {Burles} S.,  {Chen} Y.-M.,  {Dawson} K.,
  {Eisenstein} D.~J.,  {Gunn} J.~E.,  {Knapp} G.~R.,    {et al.} 2012, \aj,
  144, 144

\bibitem[\protect\citeauthoryear{{Borguet}, {Arav}, {Edmonds}, {Chamberlain} \&
  {Benn}}{{Borguet} et~al.}{2013}]{Borguet13}
{Borguet} B.~C.~J.,  {Arav} N.,  {Edmonds} D.,  {Chamberlain} C.,    {Benn} C.,
   2013, \apj, 762, 49

\bibitem[\protect\citeauthoryear{{Borguet}, {Edmonds}, {Arav}, {Benn} \&
  {Chamberlain}}{{Borguet} et~al.}{2012}]{Borguet12}
{Borguet} B.~C.~J.,  {Edmonds} D.,  {Arav} N.,  {Benn} C.,    {Chamberlain} C.,
   2012, \apj, 758, 69

\bibitem[\protect\citeauthoryear{{Brotherton}, {De Breuck} \&
  {Schaefer}}{{Brotherton} et~al.}{2006}]{Brotherton06}
{Brotherton} M.~S.,  {De Breuck} C.,    {Schaefer} J.~J.,  2006, \mnras, 372,
  L58

\bibitem[\protect\citeauthoryear{{Brotherton}, {Tran}, {Becker}, {Gregg},
  {Laurent-Muehleisen} \& {White}}{{Brotherton} et~al.}{2001}]{Brotherton01}
{Brotherton} M.~S.,  {Tran} H.~D.,  {Becker} R.~H.,  {Gregg} M.~D.,
  {Laurent-Muehleisen} S.~A.,    {White} R.~L.,  2001, \apj, 546, 775

\bibitem[\protect\citeauthoryear{{Cameron}}{{Cameron}}{2011}]{Cameron11}
{Cameron} E.,  2011, \pasa, 28, 128

\bibitem[\protect\citeauthoryear{{Capellupo}, {Hamann} \& {Barlow}}{{Capellupo}
  et~al.}{2014}]{Capellupo14}
{Capellupo} D.~M.,  {Hamann} F.,    {Barlow} T.~A.,  2014, \mnras, 444, 1893

\bibitem[\protect\citeauthoryear{{Capellupo}, {Hamann}, {Shields}, {Halpern} \&
  {Barlow}}{{Capellupo} et~al.}{2013}]{Capellupo13}
{Capellupo} D.~M.,  {Hamann} F.,  {Shields} J.~C.,  {Halpern} J.~P.,
  {Barlow} T.~A.,  2013, \mnras, 429, 1872

\bibitem[\protect\citeauthoryear{{Capellupo}, {Hamann}, {Shields},
  {Rodr{\'{\i}}guez Hidalgo} \& {Barlow}}{{Capellupo}
  et~al.}{2011}]{Capellupo11}
{Capellupo} D.~M.,  {Hamann} F.,  {Shields} J.~C.,  {Rodr{\'{\i}}guez Hidalgo}
  P.,    {Barlow} T.~A.,  2011, MNRAS, 413, 908

\bibitem[\protect\citeauthoryear{{Chamberlain}, {Arav} \& {Benn}}{{Chamberlain}
  et~al.}{2015}]{Chamberlain15}
{Chamberlain} C.,  {Arav} N.,    {Benn} C.,  2015, \mnras, 450, 1085

\bibitem[\protect\citeauthoryear{{Dawson}, {Schlegel} \& {et al.}}{{Dawson}
  et~al.}{2013}]{Dawson13}
{Dawson} K.~S.,  {Schlegel} D.~J.,    {et al.} 2013, \aj, 145, 10

\bibitem[\protect\citeauthoryear{{Dunn}, {Arav}, {Aoki}, {Wilkins}, {Laughlin},
  {Edmonds} \& {Bautista}}{{Dunn} et~al.}{2012}]{Dunn12}
{Dunn} J.~P.,  {Arav} N.,  {Aoki} K.,  {Wilkins} A.,  {Laughlin} C.,  {Edmonds}
  D.,    {Bautista} M.,  2012, \apj, 750, 143

\bibitem[\protect\citeauthoryear{{Dunn}, {Bautista}, {Arav}, {Moe}, {Korista},
  {Costantini}, {Benn}, {Ellison} \& {Edmonds}}{{Dunn} et~al.}{2010}]{Dunn10}
{Dunn} J.~P.,  {Bautista} M.,  {Arav} N.,  {Moe} M.,  {Korista} K.,
  {Costantini} E.,  {Benn} C.,  {Ellison} S.,    {Edmonds} D.,  2010, \apj,
  709, 611

\bibitem[\protect\citeauthoryear{{Eisenstein}, {Weinberg}, {Agol}, {Aihara},
  {Allende Prieto}, {Anderson}, {Arns}, {Aubourg}, {Bailey}, {Balbinot} \& et
  al.}{{Eisenstein} et~al.}{2011}]{Eisenstein11}
{Eisenstein} D.~J.,  {Weinberg} D.~H.,  {Agol} E.,  {Aihara} H.,  {Allende
  Prieto} C.,  {Anderson} S.~F.,  {Arns} J.~A.,  {Aubourg} {\'E}.,  {Bailey}
  S.,  {Balbinot} E.,    et al. 2011, \aj, 142, 72

\bibitem[\protect\citeauthoryear{{Filiz Ak}, {Brandt}, {Hall}, {Schneider},
  {Trump}, {Anderson}, {Hamann}, {Myers}, {P{\^a}ris}, {Petitjean}, {Ross},
  {Shen} \& {York}}{{Filiz Ak} et~al.}{2014}]{FilizAk14}
{Filiz Ak} N.,  {Brandt} W.~N.,  {Hall} P.~B.,  {Schneider} D.~P.,  {Trump}
  J.~R.,  {Anderson} S.~F.,  {Hamann} F.,  {Myers} A.~D.,  {P{\^a}ris} I.,
  {Petitjean} P.,  {Ross} N.~P.,  {Shen} Y.,    {York} D.,  2014, \apj, 791, 88

\bibitem[\protect\citeauthoryear{{Gabel}, {Crenshaw}, {Kraemer}, {Brandt},
  {George}, {Hamann}, {Kaiser}, {Kaspi}, {Kriss}, {Mathur}, {Mushotzky},
  {Nandra}, {Netzer}, {Peterson}, {Shields}, {Turner} \& {Zheng}}{{Gabel}
  et~al.}{2003}]{Gabel03}
{Gabel} J.~R.,  {Crenshaw} D.~M.,  {Kraemer} S.~B.,  {Brandt} W.~N.,  {George}
  I.~M.,  {Hamann} F.~W.,  {Kaiser} M.~E.,  {Kaspi} S.,  {Kriss} G.~A.,
  {Mathur} S.,  {Mushotzky} R.~F.,  {Nandra} K.,  {Netzer} H.,  {Peterson}
  B.~M.,  {Shields} J.~C.,  {Turner} T.~J.,    {Zheng} W.,  2003, \apj, 583,
  178

\bibitem[\protect\citeauthoryear{{Gibson}, {Jiang}, {Brandt}, {Hall}, {Shen},
  {Wu}, {Anderson}, {Schneider}, {Vanden Berk}, {Gallagher}, {Fan} \&
  {York}}{{Gibson} et~al.}{2009}]{Gibson09}
{Gibson} R.~R.,  {Jiang} L.,  {Brandt} W.~N.,  {Hall} P.~B.,  {Shen} Y.,  {Wu}
  J.,  {Anderson} S.~F.,  {Schneider} D.~P.,  {Vanden Berk} D.,  {Gallagher}
  S.~C.,  {Fan} X.,    {York} D.~G.,  2009, ApJ, 692, 758

\bibitem[\protect\citeauthoryear{{Glikman}, {Urrutia}, {Lacy}, {Djorgovski},
  {Mahabal}, {Myers}, {Ross}, {Petitjean}, {Ge}, {Schneider} \&
  {York}}{{Glikman} et~al.}{2012}]{Glikman12}
{Glikman} E.,  {Urrutia} T.,  {Lacy} M.,  {Djorgovski} S.~G.,  {Mahabal} A.,
  {Myers} A.~D.,  {Ross} N.~P.,  {Petitjean} P.,  {Ge} J.,  {Schneider} D.~P.,
    {York} D.~G.,  2012, \apj, 757, 51

\bibitem[\protect\citeauthoryear{{Gunn}, {Siegmund}, {Mannery}, {Owen}, {Hull},
  {Leger}, {Carey}, {Knapp}, {York}, {Boroski}, {Kent}, {Lupton} \& et
  al.}{{Gunn} et~al.}{2006}]{Gunn06}
{Gunn} J.~E.,  {Siegmund} W.~A.,  {Mannery} E.~J.,  {Owen} R.~E.,  {Hull}
  C.~L.,  {Leger} R.~F.,  {Carey} L.~N.,  {Knapp} G.~R.,  {York} D.~G.,
  {Boroski} W.~N.,  {Kent} S.~M.,  {Lupton} R.~H.,    et al. 2006, \aj, 131,
  2332

\bibitem[\protect\citeauthoryear{{Hall}, {Anosov}, {White}, {Brandt}, {Gregg},
  {Gibson}, {Becker} \& {Schneider}}{{Hall} et~al.}{2011}]{Hall11}
{Hall} P.~B.,  {Anosov} K.,  {White} R.~L.,  {Brandt} W.~N.,  {Gregg} M.~D.,
  {Gibson} R.~R.,  {Becker} R.~H.,    {Schneider} D.~P.,  2011, MNRAS, 411,
  2653

\bibitem[\protect\citeauthoryear{{Hamann}}{{Hamann}}{1998}]{Hamann98}
{Hamann} F.,  1998, ApJ, 500, 798

\bibitem[\protect\citeauthoryear{{Hamann} \& {Sabra}}{{Hamann} \&
  {Sabra}}{2004}]{Hamann04}
{Hamann} F.,  {Sabra} B.,  2004, in {Richards} G.~T.,  {Hall} P.~B.,  eds, AGN
  Physics with the Sloan Digital Sky Survey Vol.~311 of Astronomical Society of
  the Pacific Conference Series, {The Diverse Nature of Intrinsic Absorbers in
  AGNs}.
p.~203

\bibitem[\protect\citeauthoryear{{Harris}, {Jensen}, {Suzuki}, {Bautista},
  {Dawson}, {Vivek}, {Brownstein}, {Ge}, {Hamann}, {Herbst}, {Jiang}, {Moran},
  {Myers}, {Olmstead} \& {Schneider}}{{Harris} et~al.}{2016}]{Harris16}
{Harris} D.~W.,  {Jensen} T.~W.,  {Suzuki} N.,  {Bautista} J.~E.,  {Dawson}
  K.~S.,  {Vivek} M.,  {Brownstein} J.~R.,  {Ge} J.,  {Hamann} F.,  {Herbst}
  H.,  {Jiang} L.,  {Moran} S.~E.,  {Myers} A.~D.,  {Olmstead} M.~D.,
  {Schneider} D.~P.,  2016, \aj, 151, 155

\bibitem[\protect\citeauthoryear{{Hopkins} \& {Elvis}}{{Hopkins} \&
  {Elvis}}{2010}]{Hopkins10}
{Hopkins} P.~F.,  {Elvis} M.,  2010, MNRAS, 401, 7

\bibitem[\protect\citeauthoryear{{Junkkarinen}, {Beaver}, {Burbidge}, {Cohen},
  {Hamann} \& {Lyons}}{{Junkkarinen} et~al.}{1997}]{Junkkarinen97}
{Junkkarinen} V.,  {Beaver} E.~A.,  {Burbidge} E.~M.,  {Cohen} R.~D.,  {Hamann}
  F.,    {Lyons} R.~W.,  1997, in {N.~Arav, I.~Shlosman, \&amp; R.~J.~Weymann}
  ed., Mass Ejection from Active Galactic Nuclei Vol.~128 of Astronomical
  Society of the Pacific Conference Series, {On the Phosphorus Overabundance in
  the BAL QSO PG 0946+301}.
p.~220

\bibitem[\protect\citeauthoryear{{Lamy} \& {Hutsem{\'e}kers}}{{Lamy} \&
  {Hutsem{\'e}kers}}{2000}]{Lamy00}
{Lamy} H.,  {Hutsem{\'e}kers} D.,  2000, \aap, 356, L9

\bibitem[\protect\citeauthoryear{{Lamy} \& {Hutsem{\'e}kers}}{{Lamy} \&
  {Hutsem{\'e}kers}}{2004}]{Lamy04}
{Lamy} H.,  {Hutsem{\'e}kers} D.,  2004, \aap, 427, 107

\bibitem[\protect\citeauthoryear{{Leighly}, {Dietrich} \& {Barber}}{{Leighly}
  et~al.}{2011}]{Leighly11}
{Leighly} K.~M.,  {Dietrich} M.,    {Barber} S.,  2011, \apj, 728, 94

\bibitem[\protect\citeauthoryear{{Leighly}, {Hamann}, {Casebeer} \&
  {Grupe}}{{Leighly} et~al.}{2009}]{Leighly09}
{Leighly} K.~M.,  {Hamann} F.,  {Casebeer} D.~A.,    {Grupe} D.,  2009, ApJ,
  701, 176

\bibitem[\protect\citeauthoryear{{Luo}, {Brandt}, {Alexander}, {Stern}, {Teng},
  {Ar{\'e}valo}, {Bauer}, {Boggs}, {Christensen}, {Comastri} \& {et al.}}{{Luo}
  et~al.}{2014}]{Luo14}
{Luo} B.,  {Brandt} W.~N.,  {Alexander} D.~M.,  {Stern} D.,  {Teng} S.~H.,
  {Ar{\'e}valo} P.,  {Bauer} F.~E.,  {Boggs} S.~E.,  {Christensen} F.~E.,
  {Comastri} A.,    {et al.} 2014, \apj, 794, 70

\bibitem[\protect\citeauthoryear{{Misawa}, {Eracleous}, {Charlton} \&
  {Kashikawa}}{{Misawa} et~al.}{2007}]{Misawa07}
{Misawa} T.,  {Eracleous} M.,  {Charlton} J.~C.,    {Kashikawa} N.,  2007, ApJ,
  660, 152

\bibitem[\protect\citeauthoryear{{Moe}, {Arav}, {Bautista} \& {Korista}}{{Moe}
  et~al.}{2009}]{Moe09}
{Moe} M.,  {Arav} N.,  {Bautista} M.~A.,    {Korista} K.~T.,  2009, \apj, 706,
  525

\bibitem[\protect\citeauthoryear{{Ogle}, {Cohen}, {Miller}, {Tran}, {Goodrich}
  \& {Martel}}{{Ogle} et~al.}{1999}]{Ogle99}
{Ogle} P.~M.,  {Cohen} M.~H.,  {Miller} J.~S.,  {Tran} H.~D.,  {Goodrich}
  R.~W.,    {Martel} A.~R.,  1999, \apjs, 125, 1

\bibitem[\protect\citeauthoryear{{P{\^a}ris}, {Petitjean}, {Aubourg}, {Bailey},
  {Ross}, {Myers} \& {et al.}}{{P{\^a}ris} et~al.}{2012}]{Paris12}
{P{\^a}ris} I.,  {Petitjean} P.,  {Aubourg} {\'E}.,  {Bailey} S.,  {Ross}
  N.~P.,  {Myers} A.~D.,    {et al.} 2012, \aap, 548, A66

\bibitem[\protect\citeauthoryear{{P{\^a}ris}, {Petitjean}, {Rollinde},
  {Aubourg}, {Busca}, {Charlassier}, {Delubac}, {Hamilton}, {Le Goff},
  {Palanque-Delabrouille}, {Peirani}, {Pichon}, {Rich}, {Vargas-Maga{\~n}a} \&
  {Y{\`e}che}}{{P{\^a}ris} et~al.}{2011}]{Paris11}
{P{\^a}ris} I.,  {Petitjean} P.,  {Rollinde} E.,  {Aubourg} E.,  {Busca} N.,
  {Charlassier} R.,  {Delubac} T.,  {Hamilton} J.-C.,  {Le Goff} J.-M.,
  {Palanque-Delabrouille} N.,  {Peirani} S.,  {Pichon} C.,  {Rich} J.,
  {Vargas-Maga{\~n}a} M.,    {Y{\`e}che} C.,  2011, \aap, 530, A50

\bibitem[\protect\citeauthoryear{{Prochaska} \& {Hennawi}}{{Prochaska} \&
  {Hennawi}}{2009}]{Prochaska09}
{Prochaska} J.~X.,  {Hennawi} J.~F.,  2009, ApJ, 690, 1558

\bibitem[\protect\citeauthoryear{{Reichard}, {Richards}, {Schneider}, {Hall},
  {Tolea}, {Krolik}, {Tsvetanov}, {Vanden Berk}, {York}, {Knapp}, {Gunn} \&
  {Brinkmann}}{{Reichard} et~al.}{2003}]{Reichard03a}
{Reichard} T.~A.,  {Richards} G.~T.,  {Schneider} D.~P.,  {Hall} P.~B.,
  {Tolea} A.,  {Krolik} J.~H.,  {Tsvetanov} Z.,  {Vanden Berk} D.~E.,  {York}
  D.~G.,  {Knapp} G.~R.,  {Gunn} J.~E.,    {Brinkmann} J.,  2003, AJ, 125, 1711

\bibitem[\protect\citeauthoryear{{Rodr{\'{\i}}guez Hidalgo}, {Eracleous},
  {Charlton}, {Hamann}, {Murphy} \& {Nestor}}{{Rodr{\'{\i}}guez Hidalgo}
  et~al.}{2013}]{RodriguezH13}
{Rodr{\'{\i}}guez Hidalgo} P.,  {Eracleous} M.,  {Charlton} J.,  {Hamann} F.,
  {Murphy} M.~T.,    {Nestor} D.,  2013, \apj, 775, 14

\bibitem[\protect\citeauthoryear{{Rodr{\'{\i}}guez Hidalgo}, {Hamann} \&
  {Hall}}{{Rodr{\'{\i}}guez Hidalgo} et~al.}{2011}]{RodriguezH11}
{Rodr{\'{\i}}guez Hidalgo} P.,  {Hamann} F.,    {Hall} P.,  2011, MNRAS, 411,
  247

\bibitem[\protect\citeauthoryear{{Ross}, {Myers}, {Sheldon}, {Y{\`e}che},
  {Strauss}, {Bovy}, {Kirkpatrick}, {Richards}, {Aubourg}, {Blanton}, {Brandt},
  {Carithers} \& et al.}{{Ross} et~al.}{2012}]{Ross12}
{Ross} N.~P.,  {Myers} A.~D.,  {Sheldon} E.~S.,  {Y{\`e}che} C.,  {Strauss}
  M.~A.,  {Bovy} J.,  {Kirkpatrick} J.~A.,  {Richards} G.~T.,  {Aubourg}
  {\'E}.,  {Blanton} M.~R.,  {Brandt} W.~N.,  {Carithers} W.~C.,    et al.
  2012, \apjs, 199, 3

\bibitem[\protect\citeauthoryear{{Scannapieco} \& {Oh}}{{Scannapieco} \&
  {Oh}}{2004}]{Scannapieco04}
{Scannapieco} E.,  {Oh} S.~P.,  2004, ApJ, 608, 62

\bibitem[\protect\citeauthoryear{{Smee}, {Gunn}, {Uomoto}, {Roe}, {Schlegel},
  {Rockosi}, {Carr}, {Leger}, {Dawson}, {Olmstead} \& {et al.}}{{Smee}
  et~al.}{2013}]{Smee13}
{Smee} S.~A.,  {Gunn} J.~E.,  {Uomoto} A.,  {Roe} N.,  {Schlegel} D.,
  {Rockosi} C.~M.,  {Carr} M.~A.,  {Leger} F.,  {Dawson} K.~S.,  {Olmstead}
  M.~D.,    {et al.} 2013, \aj, 146, 32

\bibitem[\protect\citeauthoryear{{Sprayberry} \& {Foltz}}{{Sprayberry} \&
  {Foltz}}{1992}]{Sprayberry92}
{Sprayberry} D.,  {Foltz} C.~B.,  1992, \apj, 390, 39

\bibitem[\protect\citeauthoryear{{Trump}, {Hall}, {Reichard}, {Richards},
  {Schneider}, {Vanden Berk}, {Knapp}, {Anderson}, {Fan}, {Brinkman},
  {Kleinman} \& {Nitta}}{{Trump} et~al.}{2006}]{Trump06}
{Trump} J.~R.,  {Hall} P.~B.,  {Reichard} T.~A.,  {Richards} G.~T.,
  {Schneider} D.~P.,  {Vanden Berk} D.~E.,  {Knapp} G.~R.,  {Anderson} S.~F.,
  {Fan} X.,  {Brinkman} J.,  {Kleinman} S.~J.,    {Nitta} A.,  2006, ApJS, 165,
  1

\bibitem[\protect\citeauthoryear{{Turnshek}}{{Turnshek}}{1988}]{Turnshek88}
{Turnshek} D.~A.,  1988, in {Blades} J.~C.,  {Turnshek} D.~A.,   {Norman}
  C.~A.,  eds, Proceedings of the QSO Absorption Line Meeting {BAL QSOs -
  Observations, models and implications for narrow absorption line systems}.
pp 17--46

\bibitem[\protect\citeauthoryear{{Urrutia}, {Becker}, {White}, {Glikman},
  {Lacy}, {Hodge} \& {Gregg}}{{Urrutia} et~al.}{2009}]{Urrutia09}
{Urrutia} T.,  {Becker} R.~H.,  {White} R.~L.,  {Glikman} E.,  {Lacy} M.,
  {Hodge} J.,    {Gregg} M.~D.,  2009, \apj, 698, 1095

\bibitem[\protect\citeauthoryear{{Weymann}, {Morris}, {Foltz} \&
  {Hewett}}{{Weymann} et~al.}{1991}]{Weymann91}
{Weymann} R.~J.,  {Morris} S.~L.,  {Foltz} C.~B.,    {Hewett} P.~C.,  1991,
  ApJ, 373, 23

\bibitem[\protect\citeauthoryear{{Wright}, {Eisenhardt}, {Mainzer}, {Ressler}
  \& {et al.}}{{Wright} et~al.}{2010}]{Wright10}
{Wright} E.~L.,  {Eisenhardt} P.~R.~M.,  {Mainzer} A.~K.,  {Ressler} M.~E.,
  {et al.} 2010, \aj, 140, 1868

\bibitem[\protect\citeauthoryear{{York}, {Adelman}, {Anderson} Jr., {Anderson},
  {Annis}, {Bahcall}, {Bakken}, {Barkhouser}, {Bastian}, {Berman} \& {et
  al.}}{{York} et~al.}{2000}]{York00}
{York} D.~G.,  {Adelman} J.,  {Anderson} Jr. J.~E.,  {Anderson} S.~F.,  {Annis}
  J.,  {Bahcall} N.~A.,  {Bakken} J.~A.,  {Barkhouser} R.,  {Bastian} S.,
  {Berman} E.,    {et al.} 2000, \aj, 120, 1579

\end{thebibliography}

\appendix

\section{List and Properties of `Probable' Detections}

\begin{table*}
\caption{Properties of quasars with `probable' \pv\ detections}
\begin{tabular}{ccccccc}
\hline
SDSS Coord. Name & $z_{em}$ & $i_{mag}$ & BI & REW(\civ) & REW(\siiv) & REW(\al) \\
  &  &  & (\kms) & (\AA) & (\AA) & (\AA) \\
\hline
J001610.79$+$013608.0	& 2.839 	& 19.7 & 1363$\pm$73 	& 	& 	& 	 \\ 
J001824.95$+$001525.8	& 2.443 	& 19.4 & 7139$\pm$118 	& 46.9	& 17.1	& 0.0 \\ 
J002417.61$+$000846.2	& 4.000 	& 20.2 & 4179$\pm$245 	& 0.0	& 22.8	& 8.9 \\ 
J003859.34$-$004252.2	& 2.495 	& 20.1 & 9393$\pm$84 	& 49.4	& 30.9	& 0.0 \\ 
J003937.53$+$043025.6	& 2.819 	& 20.2 & 4752$\pm$277 	& 19.1	& 24.7	& 9.4 \\ 
J004429.13$-$015601.1	& 2.375 	& 19.6 & 2981$\pm$39 	& 21.9	& 6.7	& 0.0 \\ 
J005125.63$+$023923.4	& 3.050 	& 21.0 & 1381$\pm$328 	& 	& 	& 	 \\ 
J005708.72$+$032251.1	& 2.928 	& 20.6 & 4841$\pm$1520 	& 	& 	& 	 \\ 
J010124.98$+$023949.4	& 2.619 	& 21.1 & 1889$\pm$135 	& 	& 	& 	 \\ 
J011124.65$+$084442.2	& 2.641 	& 20.7 & 1140$\pm$301 	& 	& 	& 	 \\ 
J011301.52$-$015752.6	& 3.095 	& 20.5 & 1911$\pm$238 	& 	& 	& 	 \\ 
J013413.22$-$023409.7	& 2.394 	& 19.4 & 5495$\pm$59 	& 36.0	& 7.9	& 0.0 \\ 
J013442.35$+$001452.4	& 3.207 	& 20.6 & 3167$\pm$48 	& 21.8	& 8.9	& 0.0 \\ 
J014141.32$+$011205.7	& 3.157 	& 20.6 & 1715$\pm$74 	& 23.5	& 15.2	& 0.0 \\ 
J015949.48$+$063932.1	& 2.865 	& 20.9 & 258$\pm$26 	& 	& 	& 	 \\ 
J022007.64$-$010731.1	& 3.423 	& 18.2 & 5641$\pm$6 	& 47.7	& 8.1	& 0.0 \\ 
J073656.27$+$440308.7	& 2.700 	& 19.9 & 1232$\pm$8 	& 10.3	& 0.0	& 0.0 \\ 
J073751.52$+$455140.5	& 2.390 	& 19.6 & 3809$\pm$36 	& 27.2	& 16.3	& 0.0 \\ 
J074734.15$+$152153.0	& 2.395 	& 20.5 & 1359$\pm$32 	& 	& 	& 	 \\ 
J075014.40$+$432635.2	& 3.173 	& 19.2 & 5743$\pm$64 	& 34.0	& 13.6	& 0.0 \\ 
J080343.77$+$462345.9	& 2.458 	& 20.3 & 3758$\pm$88 	& 	& 	& 	 \\ 
J080904.19$+$505543.9	& 3.118 	& 20.8 & 3904$\pm$474 	& 	& 	& 	 \\ 
J081208.62$+$534800.3	& 2.590 	& 19.2 & 3321$\pm$28 	& 22.2	& 0.0	& 0.0 \\ 
J081410.14$+$323225.1	& 3.602 	& 20.2 & 4445$\pm$150 	& 29.3	& 3.9	& 0.0 \\ 
J081454.35$+$422453.4	& 2.350 	& 19.9 & 1686$\pm$12 	& 	& 	& 	 \\ 
J081508.91$+$122401.2	& 2.798 	& 20.9 & 3101$\pm$248 	& 	& 	& 	 \\ 
J081608.28$+$210213.2	& 3.019 	& 18.8 & 5536$\pm$50 	& 33.5	& 16.3	& 0.0 \\ 
J082227.60$+$404153.9	& 2.955 	& 20.6 & 6899$\pm$670 	& 39.9	& 17.3	& 0.0 \\ 
J082249.76$+$322712.3	& 2.495 	& 18.9 & 2606$\pm$6 	& 21.3	& 9.7	& 0.0 \\ 
J082543.23$+$383829.2	& 2.996 	& 20.1 & 3362$\pm$58 	& 20.2	& 12.6	& 0.0 \\ 
J083120.08$+$355833.8	& 3.079 	& 20.9 & 1114$\pm$40 	& 	& 	& 	 \\ 
J083715.90$+$001521.6	& 2.449 	& 19.0 & 6013$\pm$28 	& 38.3	& 17.8	& 0.0 \\ 
J090035.30$+$040846.4	& 2.809 	& 20.3 & 4250$\pm$92 	& 23.8	& 15.9	& 0.0 \\ 
J090658.65$+$021729.2	& 2.655 	& 20.7 & 4178$\pm$134 	& 	& 	& 	 \\ 
J093442.60$-$001648.9	& 2.883 	& 20.7 & 3300$\pm$358 	& 	& 	& 	 \\ 
J093707.90$-$001041.8	& 2.531 	& 19.9 & 3459$\pm$245 	& 	& 	& 	 \\ 
J094431.33$+$033411.6	& 3.006 	& 19.5 & 4324$\pm$89 	& 27.7	& 19.0	& 0.0 \\ 
J094633.97$+$365516.8	& 2.855 	& 19.1 & 2721$\pm$139 	& 11.6	& 18.9	& 0.0 \\ 
J094906.04$+$011249.9	& 2.743 	& 19.1 & 3110$\pm$27 	& 20.5	& 15.1	& 0.0 \\ 
J095333.70$+$033623.7	& 3.268 	& 19.8 & 6773$\pm$330 	& 38.6	& 32.7	& 0.0 \\ 
J095442.89$+$432512.0	& 2.478 	& 19.3 & 14542$\pm$140 	& 40.0	& 37.0	& 33.2 \\ 
J095508.25$+$014627.6	& 2.607 	& 19.3 & 3575$\pm$16 	& 27.6	& 12.3	& 0.0 \\ 
J100047.38$+$050203.7	& 2.645 	& 20.6 & 3441$\pm$206 	& 	& 	& 	 \\ 
J100049.54$-$005118.2	& 2.935 	& 19.5 & 2021$\pm$33 	& 25.0	& 2.6	& 0.0 \\ 
J101412.56$+$394135.7	& 3.269 	& 19.7 & 4880$\pm$151 	& 34.6	& 23.4	& 0.0 \\ 
J102251.29$+$031529.4	& 3.583 	& 20.6 & 2167$\pm$510 	& 	& 	& 	 \\ 
J104059.79$+$055524.4	& 2.454 	& 19.3 & 5038$\pm$58 	& 12.3	& 25.6	& 11.2 \\ 
J104247.56$+$061521.4	& 2.523 	& 20.6 & 1935$\pm$50 	& 	& 	& 	 \\ 
J105111.79$+$420356.8	& 2.932 	& 20.1 & 9993$\pm$519 	& 48.1	& 25.8	& 0.0 \\ 
J111541.03$+$335202.6	& 2.889 	& 20.4 & 3183$\pm$423 	& 	& 	& 	 \\ 
J112548.79$+$004547.1	& 3.264 	& 21.1 & 1820$\pm$39 	& 	& 	& 	 \\ 
J114548.38$+$393746.6	& 3.119 	& 17.7 & 2520$\pm$3 	& 21.2	& 9.3	& 0.0 \\ 
J114926.79$+$321902.6	& 3.228 	& 20.0 & 7222$\pm$352 	& 39.7	& 23.3	& 3.0 \\ 
J120447.15$+$330938.7	& 3.610 	& 18.4 & 13721$\pm$18 	& 51.9	& 64.5	& 42.3 \\ 
J120704.75$+$033243.9	& 2.713 	& 20.4 & 1547$\pm$56 	& 	& 	& 	 \\ 
J120834.84$+$002047.7	& 2.683 	& 18.3 & 417$\pm$2 	& 	& 	& 	 \\ 
J121858.15$+$005053.7	& 3.120 	& 20.1 & 6801$\pm$185 	& 40.1	& 21.0	& 0.0 \\ 
J122254.15$+$061041.3	& 2.441 	& 19.7 & 8786$\pm$128 	& 49.1	& 20.9	& 0.0 \\ 
J124526.35$+$341956.5	& 3.588 	& 20.7 & 525$\pm$12 	& 	& 	& 	 \\ 
J130101.92$+$382654.2	& 2.813 	& 20.0 & 4187$\pm$649 	& 	& 	& 	 \\ 
J131037.33$+$062347.8	& 3.040 	& 19.3 & 2053$\pm$46 	& 7.4	& 11.6	& 3.2 \\ 
J131333.01$-$005114.3	& 2.940 	& 19.2 & 3710$\pm$139 	& 31.4	& 22.0	& 7.4 \\ 
\end{tabular}
\label{tab:prob}
\end{table*}
\addtocounter{table}{-1}
\begin{table*}
\caption{continued...}
\begin{tabular}{ccccccc}
\hline
SDSS Coord. Name & $z_{em}$ & $i_{mag}$ & BI & REW(\civ) & REW(\siiv) & REW(\al) \\
  &  &  & (\kms) & (\AA) & (\AA) & (\AA) \\
\hline
J132004.70$+$363830.1	& 2.774 	& 19.3 & 4663$\pm$140 	& 36.6	& 20.7	& 0.0 \\ 
J132139.86$-$004151.9	& 3.091 	& 18.7 & 4755$\pm$39 	& 31.9	& 36.9	& 5.1 \\ 
J134504.32$+$071349.0	& 3.048 	& 21.3 & 397$\pm$103 	& 	& 	& 	 \\ 
J140105.31$+$062917.8	& 2.383 	& 19.4 & 4897$\pm$151 	& 34.8	& 16.2	& 9.9 \\ 
J140453.03$+$035544.9	& 2.800 	& 19.1 & 2190$\pm$85 	& 19.6	& 9.6	& 0.0 \\ 
J140532.90$+$022957.3	& 2.832 	& 18.2 & 5903$\pm$19 	& 39.3	& 12.6	& 0.0 \\ 
J141017.57$-$010657.7	& 2.817 	& 20.5 & 4832$\pm$75 	& 30.0	& 19.2	& 0.0 \\ 
J141934.64$+$050327.1	& 2.496 	& 19.4 & 3526$\pm$22 	& 20.8	& 10.7	& 0.0 \\ 
J145252.35$+$061827.4	& 3.096 	& 19.8 & 4462$\pm$113 	& 27.3	& 8.9	& 0.0 \\ 
J151102.00$+$012659.0	& 2.924 	& 19.1 & 4563$\pm$67 	& 29.6	& 15.8	& 2.2 \\ 
J151211.22$+$012807.1	& 2.705 	& 19.7 & 2673$\pm$33 	& 8.3	& 15.2	& 6.7 \\ 
J153124.37$+$213305.4	& 2.972 	& 19.7 & 11782$\pm$387 	& 38.4	& 27.7	& 17.7 \\ 
J153252.96$+$023217.2	& 2.712 	& 19.3 & 3225$\pm$55 	& 38.7	& 21.8	& 3.9 \\ 
J153637.33$+$060631.1	& 2.475 	& 20.6 & 2849$\pm$210 	& 	& 	& 	 \\ 
J154435.61$-$001928.0	& 2.659 	& 19.1 & 1851$\pm$28 	& 17.0	& 8.0	& 0.0 \\ 
J155514.85$+$100351.3	& 3.505 	& 18.3 & 8405$\pm$19 	& 36.9	& 30.7	& 24.0 \\ 
J161626.54$+$121955.7	& 2.975 	& 19.6 & 5610$\pm$110 	& 23.3	& 41.1	& 18.2 \\ 
J163402.87$+$251635.4	& 2.602 	& 19.7 & 1794$\pm$8 	& 12.7	& 5.7	& 0.0 \\ 
J165905.52$+$342937.5	& 2.531 	& 20.5 & 2765$\pm$170 	& 	& 	& 	 \\ 
J211137.38$-$023941.9	& 3.180 	& 18.4 & 6083$\pm$44 	& 37.0	& 31.7	& 3.8 \\ 
J214244.85$+$004528.2	& 2.498 	& 19.2 & 3437$\pm$12 	& 23.3	& 11.0	& 0.0 \\ 
J222725.45$-$000936.1	& 2.566 	& 19.6 & 5604$\pm$19 	& 32.2	& 20.1	& 3.2 \\ 
J231656.44$-$002240.9	& 3.057 	& 21.0 & 3314$\pm$615 	& 	& 	& 	 \\ 
J235859.51$+$020847.5	& 2.921 	& 19.0 & 5389$\pm$83 	& 31.4	& 12.6	& 0.0 \\  
\end{tabular}
\end{table*}

\bsp

\label{lastpage}

\end{document}